\documentclass[twocolumn,showpacs,preprintnumbers,amsmath,amssymb]{revtex4}

\usepackage{graphicx}
\usepackage{dcolumn}
\usepackage{bm}

\begin{document}

\preprint{APS/123-QED}

\title{Influence of helicity on anomalous scaling of a passive scalar advected by the turbulent
velocity field with finite correlation time: \\ Two-loop
approximation}

\author{O.G.\,Chkhetiani,$^1$ M.\,Hnatich,$^{2,3}$ E.\,Jur\v{c}i\v{s}inov\'a,$^{2,4}$
        M.\,Jur\v{c}i\v{s}in,$^{2,5}$ A.\,Mazzino,$^{6}$ and M.\,Repa\v{s}an$^{2}$}
\address{$^{1}$ Space Research Institute, Profsoyuznaya 84/32, 117 997 Moscow, Russian Federation \\
         $^{2}$ Institute of Experimental Physics, Slovak Academy of Sciences,
                Watsonova 47, 043 53 Ko\v{s}ice, Slovakia \\
         $^{3}$ Department of Mathematics Faculty of Civil Engineering, Technical
         University, Vysoko\v{s}kolsk\'a 4, 040 01 Ko\v{s}ice, Slovakia
                \\
         $^{4}$ Laboratory of Information Technologies, Joint Institute for Nuclear Research,
                141 980 Dubna, Moscow Region, Russian Federation \\
         $^{5}$ N.N. Bogoliubov Laboratory of Theoretical Physics, Joint Institute for
                Nuclear Research, 141 980 Dubna, Moscow Region, Russian Federation \\
         $^{6}$ Dipartamento di Fisica, Universit$\grave{a}$ di
                Genova, I-16146, Genova, Italy
        } \draft

\date{\today}

\begin{abstract}
The influence of helicity on the stability of scaling regimes, on
the effective diffusivity, and on the anomalous scaling of structure
functions of a passive scalar advected by a Gaussian solenoidal
velocity field with finite correlation time is investigated by the
field theoretic renormalization group and operator product expansion
within the two-loop approximation. The influence of helicity on the
scaling regimes is discussed and shown in the plane of exponents
$\varepsilon-\eta$, where $\varepsilon$ characterizes the energy
spectrum of the velocity field in the inertial range $E\propto
k^{1-2\varepsilon}$, and $\eta$ is related to the correlation time
at the wave number $k$, which is scaled as $k^{-2+\eta}$. The
restrictions given by nonzero helicity on the regions with stable
fixed points that correspond to the scaling regimes are analyzed in
detail. The dependence of the effective diffusivity on the helicity
parameter is discussed. The anomalous exponents of the structure
functions of the passive scalar field which define their anomalous
scaling are calculated and it is shown that, although the separate
composite operators which define them strongly depend on the
helicity parameter, the resulting two-loop contributions to the
critical dimensions of the structure functions are independent of
helicity. Details of calculations are shown.
\end{abstract}

\maketitle

\section{\label{sec:level1}Introduction}

During the last decade much attention has been paid to the inertial
range of fully developed turbulence, which contains wave numbers
larger than those that pump the energy into the system and smaller
than those that are related to the dissipation processes
\cite{MonYagBook,HuPhWi91}. Grounding for the inertial range
turbulence was created in the well known
 Kolmogorov--Obukhov (KO) phenomenological theory
(see, e.g., \cite{MonYagBook,McComb,Frisch}). One of the main
problems in the modern theory of fully developed turbulence is to
verify the validity of the basic principles of  KO theory and their
consequences within the framework of a microscopic model. Recent
experimental and theoretical studies indicate  possible deviations
from the celebrated Kolmogorov scaling exponents. The scaling
behavior  of velocity fluctuations with exponents whose values are
different from Kolmogorov ones is called anomalous and usually is
associated with the intermittency phenomenon. In turbulence this
phenomenon is believed to be related to strong fluctuations of the
energy flux which therefore leads to the deviations from the
predictions of the aforementioned KO theory. Such deviations,
referred to as anomalous or nondimensional scaling, manifest
themselves in a singular dependence of the correlation or structure
functions on the distances and the integral (external) turbulence
scale $L$. The corresponding exponents are certain nontrivial and
nonlinear functions of the order of the correlation function, the
phenomenon referred to as ``multiscaling.'' Even though great
progress in the understanding of intermittency and anomalous scaling
in turbulence was achieved as a result of intensive studies, their
investigation in fully developed turbulence still remains a major
theoretical problem.

Although the theoretical description of fluid turbulence on the
basis of first principles, i.e., on the stochastic Navier-Stokes
equation \cite{MonYagBook}, remains essentially an open problem,
considerable progress has been achieved in understanding simplified
model systems that share some important properties with the real
problem: shell models \cite{Dyn}, the stochastic Burgers equation
\cite{Burgulence}, and passive advection by random ``synthetic''
velocity fields \cite{FaGaVe01}.

A crucial role in these studies is played by models of the advected
passive scalar field \cite{Obu49}. A simple model of a passive
scalar quantity advected by a random Gaussian velocity field, white
in time and self-similar in space (the latter property mimics some
features of a real turbulent velocity ensemble), the so-called
Kraichnan rapid-change model \cite{Kra68}, is an example. The
interest in these models is based on two important facts: first, as
shown by both natural and numerical experimental investigations, the
deviations from the predictions of the classical Kolmogorov-Obukhov
phenomenological theory \cite{MonYagBook,OrszagBook,Frisch,McComb}
are even more strongly displayed for a passively advected scalar
field than for the velocity field itself (see, e.g.,
\cite{AnHoGaAn84,Sre91,HolSig94,Pumir94,TonWar94,ElKlRo90te} and
references cited therein), and second, the problem of passive
advection is much easier to consider from a theoretical point of
view. In these studies the anomalous scaling was established on the
basis of a microscopic model \cite{Kraichnan94}, and corresponding
anomalous exponents were calculated within controlled approximations
\cite{all,Shraiman,Pumir} (see also the review \cite{FaGaVe01} and
references therein).

The greatest stimulation to study the simple models of passive
advection not only of scalar fields but also of vector fields (e.g.,
a weak magnetic field) is related to the fact that even simplified
models with given Gaussian statistics of a so-called synthetic
velocity field describes a lot of features of the anomalous behavior
of genuine turbulent transport of some quantities (such as heat or
mass) observed in experiments (see, e.g.,
Refs.\,\cite{Kra68,HolSig94,Pumir94,TonWar94,ElKlRo90te,all,Shraiman,Pumir,AveMaj90,ZhaGli92,Kraichnan94,
KrYaCh95} and references cited therein).

An effective method for investigation of a self-similar scaling
behavior is the renormalization group (RG) technique
\cite{ZinnJustin,Vasiliev,Collins}. It was widely used in the theory
of critical phenomena to explain the origin of the critical scaling
and also to calculate corresponding universal quantities (e.g.,
critical dimensions). This method can also be directly used in the
theory of turbulence
\cite{Vasiliev,deDoMa79,AdVaPi83,AdAnVa96,AdAnVa99}, as well as in
related models like the simpler stochastic problem of a passive
scalar advected by a prescribed stochastic flow. In what follows we
use the conventional ("quantum field theory" or field theoretic) RG
which is based on the standard renormalization procedure, i.e., on
elimination of the ultraviolet (uv) divergences.

In Refs.\,\cite{AdAnVa98+} the field theoretic RG and
operator-product expansion (OPE) were used in the systematic
investigation of the rapid-change model. It was shown that within
the field theoretic approach the anomalous scaling is related to the
very existence of so-called dangerous composite operators with
negative critical dimensions in the OPE (see, e.g.,
\cite{Vasiliev,AdAnVa99} for details). In subsequent papers
\cite{AdAnBaKaVa01} the anomalous exponents of the model were
calculated within the $\varepsilon$ expansion to order
$\varepsilon^3$ (three-loop approximation). Here $\varepsilon$ is a
parameter that describes a given equal-time pair correlation
function of the velocity field (see later section). Important
advantages of the RG approach are its universality and calculational
efficiency: a regular systematic perturbation expansion for the
anomalous exponents was constructed, similar to the well-known
$\epsilon$ expansion in the theory of phase transitions.

Afterward, various generalized descendants of the Kraichnan model,
namely, models with inclusion of large and small scale anisotropy
\cite{AdAnHnNo00}, compressibility \cite{AdAn98}, and finite
correlation time of the velocity field \cite{Antonov99,Antonov00}
were studied by the field theoretic approach. Moreover, advection of
a passive vector field by the Gaussian self-similar velocity field
(with and without large and small scale anisotropy, pressure,
compressibility, and finite correlation time) has also been
investigated and all possible asymptotic scaling regimes and
crossovers among them have been classified \cite{all1}. The general
conclusion is that the anomalous scaling, which is the most
important feature of the Kraichnan rapid-change model, remains valid
for all generalized models.

Let us describe briefly the solution of the problem in the framework
of the field theoretic approach (see, e.g.,
Refs.\,\cite{Vasiliev,AdAnVa96,AdAnVa99} for more details). It can
be divided into two main stages. In the first stage the
multiplicative renormalizability of the corresponding field
theoretic model is demonstrated and the differential RG equations
for its correlation functions are obtained. The asymptotic behavior
of the latter on their ultraviolet argument $(r/\ell)$ for
$r\gg\ell$ and any fixed $(r/L)$ is given by infrared  stable fixed
points of those equations. Here $\ell$ and $L$ are the inner
(ultraviolet) and the outer (infrared) scales. The behavior involves
some {}``scaling functions'' of the infrared argument $(r/L)$, whose
form is not determined by the RG equations. In the second stage,
their behavior at $r\ll L$ is found from the OPE within the
framework of the general solution of the RG equations. There, the
crucial role is played by the critical dimensions of various
composite operators, which give rise to an infinite family of
independent scaling exponents as mentioned above (and hence to
multiscaling). Of course, both these stages (and thus the phenomenon
of multiscaling) have long been known in the RG theory of critical
behavior. The distinguishing feature specific to models of
turbulence  is the existence of composite operators with the
aforementioned {\it negative} critical dimensions. Their
contributions to the OPE diverge at $(r/L)\to0$. In models of
critical phenomena, nontrivial composite operators always have
strictly positive dimensions, so that they only determine
corrections [vanishing for $(r/L)\to0$] to the leading terms [finite
for $(r/L)\to0$] in the scaling functions.

In Ref.\,\cite{Antonov99} the problem of a passive scalar advected
by a Gaussian self-similar velocity field with finite correlation
time \cite{all2} was studied by the field theoretic RG method.
There, a systematic study of the possible scaling regimes and
anomalous behavior was presented at one-loop level. The two-loop
corrections to the anomalous exponents were obtained in
Ref.\,\cite{AdAnHo02}. It was shown that the anomalous exponents are
nonuniversal as a result of their dependence on a dimensionless
parameter, the ratio of the velocity correlation time and the
turnover time of the scalar field.

In what follows, we shall continue with the investigation of this
model from the point of view of the influence of helicity (spatial
parity violation) on the scaling regimes and anomalous exponents
within the two-loop approximation.

Helicity is defined as the scalar product of velocity and vorticity
and its nonzero value expresses mirror symmetry breaking of the
turbulent flow. It plays a significant role in the processes of
magnetic field generation in an electrically conductive fluid
\cite{dynamo1}-\cite{Steh3} and represents one of the most important
characteristics of large-scale motions as well
\cite{Etling85}-\cite{Ponomar2003}. The presence of helicity is
observed in various natural (like large air vortices in the
atmosphere) and technical flows
\cite{Moffat92,Kholmyansky2001,Koprov2005}. Despite this fact the
role of the helicity in hydrodynamical turbulence is not completely
clarified up to now.

The Navier-Stokes equations conserve kinetic energy and helicity in
the inviscid limit. The presence of two quadratic invariants leads
to the possibility of appearance of double cascade. This means that
cascades of energy and helicity take place in different ranges of
wave numbers analogously to the two-dimensional turbulence and/or
the helicity cascade  appearing  concurrently to the energy cascade
in the direction of small scales \cite{Brissaud73,Moiseev1996}. In
particular, the helicity cascade is closely connected with the
existence of the exact relation between triple and double
correlations of velocity known as the 2/15 law analogously to the
4/5 Kolmogorov law \cite{Chkhet96}. Corresponding to
\cite{Brissaud73} the aforementioned scenarios of turbulent cascades
differ from each other by spectral scaling. Theoretical arguments
given by Kraichnan \cite{Kraich73}  and results of numerical
calculations of the Navier-Stokes equations
\cite{Andre77,Orszag1997,Chen2003} support the scenario of
concurrent cascades. The appearance of helicity in turbulent systems
leads to the constraint of the nonlinear cascade to small scales.
This phenomenon was first demonstrated by Kraichnan \cite{Kraich73}
within the modeling problem of statistically equilibrium spectra and
later in numerical experiments.

Turbulent viscosity and diffusivity, which characterize the
influence of small-scale motions on heat and momentum transport, are
basic quantities investigated in  the theoretic and applied models.
The constraint of the direct energy cascade in helical turbulence
has to be accompanied by a decrease of turbulent viscosity. However,
no influence of helicity on turbulent viscosity was found in some
works \cite{Pouquet78,Zhou91}. A similar situation is observed for
the turbulent diffusivity in helical turbulence. Although the
modeling calculations demonstrate intensification of turbulent
transfer in the presence of helicity \cite{Kraich761,Drum84} direct
calculation of the diffusivity  does not confirm this effect
\cite{Kraich761,Knobl77,Lipscombe91}. Helicity is a pseudoscalar
quantity, hence, it can be easily understood that its influence
appears only in quadratic and higher terms of perturbation theory or
in combination with other pseudoscalar quantities (e.g., large-scale
helicity). Really, simultaneous consideration of memory effects and
second order approximation indicates the effective influence of
helicity on turbulent viscosity \cite{Belian1994,Belian1998} and
turbulent diffusivity
\cite{Drum84,Dolg87,Drummond2001,Chkhetiani2004} already in the
limits of small and infinite correlation time.

Helicity, as we shall see below, does not affect known results in
the one-loop approximation and,  therefore, it is necessary to turn
to the second order (two-loop) approximation to be able to analyze
possible consequences. It is also important to say that in the
framework of the classical Kraichnan model, i.e., a model of passive
advection by a Gaussian velocity field with $\delta$-like
correlations in time, it is not possible to study the influence of
the helicity because all potentially helical diagrams are
identically equal to zero at all orders in the perturbation theory.
In this sense, the investigation of the helicity in the present
model can be consider as the first step toward  analyzing  the
helicity in genuine turbulence. In fact, it is interesting and
important to study the helicity effects because many turbulence
phenomena are directly influenced by them (like large air vortices
in the atmosphere). For example, in stochastic magnetic
hydrodynamics, which studies the turbulence in electrically
conducting fluids, it leads to the  nontrivial fact of the existence
of a so-called turbulent dynamo -- the generation of a large-scale
magnetic field by the energy of the turbulent motion
\cite{dynamo1,Moffatt,AdVaHn87,HnJuSt01,Steh1,Steh2,Steh3}. This is
an important effect in astrophysics.


The main result of the paper will be the conclusion that helicity
does not change the anomalous exponents of the single-time structure
functions within the two-loop approximation although the separate
composite operators which define them strongly depend on the
helicity parameter. This result leads to the following interesting
but nontrivial question: is this result related only to the two-loop
calculations or does it hold for all orders of perturbation theory?
Of course, the answer to this question definitely lies at least
within the three-loop approximation. On the other hand, as will be
shown, the effective diffusivity rather strongly depends on the
helicity parameter.


The paper is organized as follows. In Sec.\,\ref{sec:Model} we
present the definition of the model and introduce the helicity to
the transverse projector of a given pair correlation function of the
velocity field. In Sec.\,\ref{sec:Field} we give the field theoretic
formulation of the original stochastic problem and discuss the
corresponding diagrammatic technique. In Sec.\,\ref{sec:RG} we
analyze the ultraviolet divergences of the model, establish its
multiplicative renormalizability, and calculate the renormalization
constants in the two-loop approximation. In Sec.\,\ref{sec:ScalReg}
we analyze possible scaling regimes of the model, associated with
nontrivial and physically acceptable fixed points of the
corresponding RG equations. There are five such regimes, any one of
which can be realized in dependence  on the values of the parameters
of the model. We discuss the physical meaning of these regimes
(e.g., some of them correspond to zero, finite, or infinite
correlation time of the advecting field) and their regions of
stability in the space of the model parameters. In
Sec.\,\ref{sec:EffDiff}  the two-loop corrections to the effective
diffusivity are calculated. In Sec.\,\ref{sec:CompOper} the
renormalization of needed composite operators is done and their
explicit dependence on the helicity parameter is shown.  In
Sec.\,\ref{sec:Conc} discussion of the results is present.

\section{The model} \label{sec:Model}

In what follows, we shall consider the advection of a passive scalar
field $\theta \equiv \theta(x)\equiv \theta(t, {\bf x})$ which is
described by the following stochastic equation:
\begin{equation}
\partial_t \theta + v_i \partial_i \theta=\nu_0 \Delta
\theta + f^{\theta},\label{scalar1}
\end{equation}
where $\partial_t \equiv \partial/\partial t$, $\partial_i \equiv
\partial/\partial x_i$, $\nu_0$ is the coefficient of molecular
diffusivity (hereafter all parameters with a subscript $0$ denote
bare parameters of the unrenormalized theory; see below), $\Delta
\equiv \partial^2$ is the Laplace operator, $v_i \equiv v_i(x)$ is
the $i$th component of the divergence-free (owing to the
incompressibility) velocity field ${\bf v}(x)$, and $f^{\theta}
\equiv f^{\theta}(x)$ is a Gaussian random noise with zero mean and
correlation function
\begin{equation}
\langle f^{\theta}(x) f^{\theta}(x^{\prime})\rangle =
\delta(t-t^{\prime})C({\bf r}/\tilde{L}), \,\,\, {\bf r}={\bf
x}-{\bf x^{\prime}},\label{correlator}
\end{equation}
where the angular brackets $\langle...\rangle$ hereafter denote the
average over the corresponding statistical ensemble. The noise
maintains the steady state of the system but the concrete form of
the correlator is not essential. The only condition that must be
satisfied by the function $C({\bf r}/\tilde{L})$ is that it must
decrease rapidly for $r\equiv |{\bf r}| \gg \tilde{L}$, where
$\tilde{L}$ denotes an integral scale related to the stirring. In
the case when $C$ depends not only on the modulus of the vector
${\bf r}$ but also on its direction, it plays the role of a source
of large-scale anisotropy, whereupon the noise can be replaced by a
constant gradient of scalar field. Equation (\ref{scalar1}) then
reads (see, e.g., Ref.\,\cite{Antonov99})
\begin{equation}
\partial_t \theta + v_i \partial_i \theta=\nu_0 \Delta
\theta - {\bf h}\cdot{\bf v}.\label{scalar2}
\end{equation}
Here, $\theta(x)$ is the fluctuation part of the total scalar field
$\Theta(x)=\theta(x)+{\bf h}\cdot{\bf x}$, and ${\bf h}$ is a
constant vector that determines the distinguished direction. The
direct formulation with a scalar gradient is even more realistic;
see, e.g. Refs.\,\cite{HolSig94,Shraiman,Pumir,Antonov99,Antonov00}.

In real problems the velocity field ${\bf v}(x)$ satisfies the
stochastic Navier-Stokes equation. In spite of this fact, in what
follows, we shall suppose that the velocity field is driven by the
simple linear stochastic equation \cite{HolSig94,Antonov99}
\begin{equation}
\partial_t v_i + R v_i=f^v_i, \label{linearNS}
\end{equation}
where $R\equiv R(x)$ is a linear operation to be specified below
and $f^v_i\equiv f^v_i(x)$ is an external random stirring force
with zero mean and the correlator
\begin{eqnarray}
\langle f^v_i(x) f^v_j(x^{\prime})\rangle &\equiv&  D^f_{ij}(x;
x^{\prime})\nonumber \\ &=&  \int \frac{d\omega d^d
k}{(2\pi)^{d+1}} P^{\rho}_{ij}({\bf k}) \tilde{D}^f(\omega,k) \nonumber \\
&& \times \exp[-i(t-t^{\prime})+i{\bf k}({\bf x}-{\bf x^{\prime}})],
\label{corf}
\end{eqnarray}
where $k=|{\bf k}|$ is the wave number, $\omega$ is the frequency,
$d$ is the dimensionality of the ${\bf x}$ space (of course, when
one investigates a system with helicity the dimension of the ${\bf
x}$ space must be strictly equal to 3; nevertheless, in what
follows, we shall retain the $d$-dimensionality of all results that
are not related to helicity so that we can also study the $d$
dependence of the nonhelical case of the model). The transition to a
helical fluid corresponds to the giving up of conservation of
spatial parity, and technically this is expressed by the fact that
the correlation function is specified in the form of a mixture of a
true tensor and a pseudotensor. In our approach, it is represented
by two parts of the transverse projector
\begin{equation}
P^{\rho}_{ij}=P_{ij}({\bf k})+H_{ij}({\bf k}), \label{projectorA}
\end{equation}
which consists of the nonhelical standard transverse projector
$P_{ij}({\bf k})=\delta_{ij}-k_i k_j/k^2$ and $H_{ij}({\bf k})=i
\rho \, \varepsilon_{ijl} k_l/k$, which represents the presence of
helicity in the flow. Here, $\varepsilon_{ijl}$ is Levi-Civita's
completely antisymmetric tensor of rank 3 [it is equal to $1$ or
$-1$ according to whether $(i,j,l)$ is an even or odd permutation of
$(1,2,3)$ and zero otherwise], and the real parameter of helicity,
$\rho$, characterizes the amount of helicity. Due to the requirement
of positive definiteness of the correlation function the absolute
value of $\rho$ must be in the interval $|\rho| \in \langle
0,1\rangle$ \cite{AdVaHn87,HnJuSt01}. Physically, the nonzero
helical part (proportional to $\rho$) expresses the existence of
nonzero correlations $\langle{\bf v}\cdot \mathrm{rot}\:{\bf
v}\rangle$.

We choose the correlator $D^f$ in Eq.\,(\ref{corf}) to be a $\delta$
function in time, which is equivalent to the condition that
$\tilde{D}^f$ is independent of frequency \cite{HolSig94} (see also
Refs.\cite{Antonov99,Antonov00}). Following
\cite{Antonov99,Antonov00}, we shall work with
\begin{equation}
\tilde{D}^f(\omega, k)=g_0 \nu_0^3
(k^2+m^2)^{2-d/2-\varepsilon-\eta/2}\label{Df}
\end{equation}
and
\begin{equation}
\tilde{R}(k)=u_0 \nu_0 (k^2+m^2)^{1-\eta/2},\label{R}
\end{equation}
the wave-number representation of $R(x)$. Here, the positive
amplitude factors $g_0$ and $u_0$ play the roles of the coupling
constants of the model, the analogs of the coupling constant
$\lambda_0$ in the $\lambda_0 \varphi^4$ model of critical behavior
\cite{ZinnJustin,Vasiliev}. In addition, $g_0$ is a formal small
parameter of the ordinary perturbation theory. The positive
exponents $\varepsilon$ and $\eta$ [$\varepsilon=O(\eta)$] are small
RG expansion parameters, the analogs of the parameter
$\varepsilon=4-d$ in the $\lambda_0 \varphi^4$ theory. Thus, we have
a kind of double expansion model in the $\varepsilon -\eta$ plane
around the origin $\varepsilon=\eta=0$. An integral scale $L=1/m$ is
introduced to provide infrared (ir) regularization. In the limit $k
\gg m$ the functions (\ref{Df}) and (\ref{R}) take on simple
powerlike forms
\begin{equation}
\tilde{D}^f(\omega, k)=g_0 \nu_0^3 k^{4-d-2 \varepsilon-\eta},\,\,\,
\tilde{R}(k)=u_0 \nu_0 k^{2-\eta},\label{corf1}
\end{equation}
which will be used in calculations in what follows. The needed ir
regularization will be given by restrictions on the region of
integration.

From Eqs.\,(\ref{linearNS}), (\ref{corf}), and (\ref{corf1}) we
receive the statistics of the velocity field ${\bf v}$. It obeys a
Gaussian distribution with zero mean and correlator
\begin{eqnarray}
\langle v_i(x) v_j(x^{\prime}) \rangle &\equiv&  D^v_{ij}(x;
x^{\prime})\nonumber \\ &=&  \int \frac{d\omega d^d
k}{(2\pi)^{d+1}} P^{\rho}_{ij}({\bf k}) \tilde{D}^v(\omega,k) \nonumber \\
&& \times \exp[-i\omega(t-t^{\prime})+i{\bf k}({\bf x}-{\bf
x^{\prime}})], \label{corv}
\end{eqnarray}
with
\begin{equation}
\tilde{D}^v(\omega, k) = \frac{g_0 \nu_0^3
k^{4-d-2\varepsilon-\eta}}{(i\omega+u_0 \nu_0
k^{2-\eta})(-i\omega+u_0 \nu_0 k^{2-\eta})}.\label{corrvelo}
\end{equation}
The correlator (\ref{corrvelo}) is directly related to the energy
spectrum via the frequency integral
\cite{AveMaj90,ZhaGli92,Antonov99}
\begin{equation}
E(k)\simeq k^{d-1} \int d\omega \tilde{D}^v(\omega, k) \simeq
\frac{g_0 \nu_0^2}{u_0} k^{1-2\varepsilon}.
\end{equation}
Therefore, the coupling constant $g_0$ and the exponent
$\varepsilon$ describe the equal-time velocity correlator or,
equivalently, the energy spectrum. On the other hand, the constant
$u_0$ and the second exponent $\eta$ are related to the frequency
$\omega \simeq u_0 \nu_0 k^{2-\eta}$ [or to the function
$\tilde{R}(k)$, the reciprocal of the correlation time at the wave
number $k$] which characterizes the mode $k$
\cite{AveMaj90,ZhaGli92,Antonov99,CheFalLeb96,Eyink96}. Thus, in our
notation, the value $\varepsilon=4/3$ corresponds to the well-known
Kolmogorov five-thirds law for the spatial statistics of velocity
field, and $\eta=4/3$ corresponds to the Kolmogorov frequency.
Simple dimensional analysis shows that the parameters (charges)
$g_0$ and $u_0$ are related to the characteristic ultraviolet
momentum scale $\Lambda$ (of the order of the inverse Kolmogorov
length) by
\begin{equation}
g_0\simeq \Lambda^{2\varepsilon + \eta},\,\,\, u_0\simeq
\Lambda^{\eta}.
\end{equation}

In Ref.\,\cite{HolSig94} it was shown that the linear model
(\ref{linearNS}) [and therefore also the Gaussian model
(\ref{corv}), (\ref{corrvelo})] is not Galilean invariant and, as a
consequence, it does not take into account the self-advection of
turbulent eddies. As a result of these so-called sweeping effects
the different time correlations of the Eulerian velocity are not
self-similar and depend strongly on the integral scale; see, e.g.,
Ref.\,\cite{kraichnan}. But, on the other hand, the results
presented in Ref.\,\cite{HolSig94} show that the Gaussian model
gives a reasonable description of the passive advection in the
appropriate frame, where the mean velocity field vanishes. One more
argument to justify the model (\ref{corv}), (\ref{corrvelo}) is
that, in what follows, we shall be interested in the equal-time,
Galilean-invariant quantities (structure functions), which are not
affected by the sweeping, and therefore, as we expect (see, e.g.,
Refs.\,\cite{Antonov99,Antonov00,AdAnHo02}), their absence in the
Gaussian model (\ref{corv}), (\ref{corrvelo}) is not essential.

At the end of this section, let us briefly discuss two important
limits of the considered model (\ref{corv}), (\ref{corrvelo}). The
first is the so-called  rapid-change model limit when
$u_0\rightarrow \infty$ and $g_0^{\prime}\equiv g_0/u_0^2=$ const,
\begin{equation}
\tilde{D}^v(\omega, k)\rightarrow g_0^{\prime} \nu_0
k^{-d-2\varepsilon + \eta},
\end{equation}
and the second is the so-called quenched (time-independent or
frozen) velocity field limit, which is defined by $u_0\rightarrow 0$
and $g_0^{\prime\prime}\equiv g_0/u_0=$ const,
\begin{equation}
\tilde{D}^v(\omega, k)\rightarrow g_0^{\prime\prime} \nu_0^2 \pi
\delta(\omega) k^{-d+2-2\varepsilon},
\end{equation}
which is similar to the well-known models of the random walks in
random environment with long-range correlations; see, e.g.,
Refs.\,\cite{Bouchaud,Honkonen}.

\section{\label{sec:Field}Field Theoretic Formulation of the Model}

For completeness of our text in this and the next section we shall
present and discuss  the principal moments of the RG theory of the
model defined by Eqs.\,(\ref{scalar2}), (\ref{corv}), and
(\ref{corrvelo}).

We start with the reformulation of the stochastic problem
(\ref{scalar2})-(\ref{corf}), according to the well-known general
theorem (see, e.g., Refs.\cite{ZinnJustin,Vasiliev}), into the
equivalent field theoretic model of the doubled set of fields $\Phi
\equiv \{\theta, \theta^{\prime}, {\bf v}, {\bf v^{\prime}}\}$ with
the following action functional:
\begin{eqnarray}
S(\Phi)&=&\frac{1}{2} \int dt_1\,d^d{\bf x_1}\,dt_2\,d^d{\bf x_2}
\nonumber \\ && v_i^{\prime}(t_1,{\bf x_1}) D_{ij}^f(t_1,{\bf
x_1};t_2,{\bf x_2}) v_j^{\prime}(t_2,{\bf x_2}) \label{action1} \\
&+& \int dt\,d^d{\bf x}\,\, \theta^{\prime}\left[-\partial_t \theta
- v_i\partial_i\theta+\nu_0\triangle\theta-{\bf h}\cdot{\bf v}
\right] \nonumber \\
&+& \int dt\,d^d{\bf x}\,\, v_i^{\prime}\left[-\partial_t - R
\right] v_i ,\nonumber
\end{eqnarray}
where $D_{ij}^f$ is defined in Eq.\,(\ref{corf}),
$\theta^{\prime}$ and ${\bf v^{\prime}}$ are auxiliary scalar and
vector fields, and summations are implied over the vector indices.

It is standard that the formulation through the action functional
(\ref{action1}) replaces the statistical averages of random
quantities in the stochastic problem (\ref{scalar2})-(\ref{corf})
with equivalent functional averages with weight $\exp S(\Phi)$. The
generating functionals of the total Green's functions G(A) and
connected Green's functions W(A) are then defined by the functional
integral
\begin{equation}
G(A)=e^{W(A)}=\int {\cal D}\Phi \,\, e^{S(\Phi) +
A\Phi},\label{green}
\end{equation}
where $A(x)=\{A^{\theta},A^{\theta^{\prime}},{\bf A^{v}},{\bf
A^{v^{\prime}}}\}$ represents a set of arbitrary sources for the set
of fields $\Phi$, ${\cal D}\Phi \equiv {\cal D}\theta{\cal
D}\theta^{\prime}{\cal D}{\bf v}{\cal D}{\bf v^{\prime}}$ denotes
the measure of functional integration, and the linear form $A\Phi$
is defined as
\begin{eqnarray}
A\Phi&=& \int d\,x
[A^{\theta}(x)\theta(x)+A^{\theta^{\prime}}(x)\theta^{\prime}(x)
\nonumber \\ &&+\,\, A_i^{v}(x) v_i(x)+A_i^{v^{\prime}}(x)
v_i^{\prime}(x)].\label{form}
\end{eqnarray}
Following the arguments in \cite{Antonov99}, we can put $A_i^{{\bf
v^{\prime}}}=0$ in Eq.\,(\ref{form}) and then perform an explicit
Gaussian integration over the auxiliary vector field ${\bf
v^{\prime}}$ in Eq.\,(\ref{green}) as a consequence of the fact
that, in what follows, we shall not be interested in the Green's
functions involving the field ${\bf v^{\prime}}$. After this
integration one is left with the field theoretic model described by
the functional action
\begin{eqnarray}
S(\Phi)&=&-\frac{1}{2} \int dt_1\,d^d{\bf x_1}\,dt_2\,d^d{\bf x_2}
\nonumber \\ && v_i(t_1,{\bf x_1}) [D_{ij}^v(t_1,{\bf
x_1};t_2,{\bf x_2})]^{-1} v_j(t_2,{\bf x_2}) \label{action3} \\
&+& \int dt\,d^d{\bf x}\,\, \theta^{\prime}\left[-\partial_t \theta
- v_i\partial_i\theta+\nu_0\triangle\theta-{\bf h}\cdot{\bf v}
\right],\nonumber
\end{eqnarray}
where the four terms in the third line in Eq.\,(\ref{action3})
represent the Martin-Siggia-Rose action for the stochastic problem
(\ref{scalar2}) at fixed velocity field ${\bf v}$, and the first two
lines describe the Gaussian averaging over ${\bf v}$ defined by the
correlator $D^v$ in Eqs.\,(\ref{corv}) and (\ref{corrvelo}).

\input epsf
   \begin{figure}[t]
     \vspace{0cm}
       \begin{center}
       \leavevmode
       \epsfxsize=5cm
       \epsffile{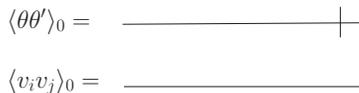}
   \end{center}
\vspace{0cm} \caption{The graphical representation of the
propagators of the model.\label{propagators}}
\end{figure}

\input epsf
   \begin{figure}[b]
     \vspace{0cm}
       \begin{center}
       \leavevmode
       \epsfxsize=5cm
       \epsffile{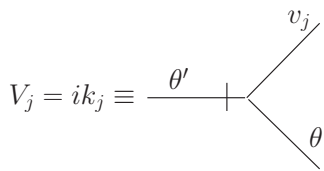}
   \end{center}
\vspace{0cm} \caption{The interaction vertex of the model
(wave-number- frequency representation). \label{vertex}}
\end{figure}

The action (\ref{action3}) is given in a form convenient for a
realization of the field theoretic perturbation analysis with the
standard Feynman diagrammatic technique. From the quadratic part of
the action one obtains the matrix of bare propagators. The
wave-number-frequency representation of, in what follows, important
propagators is as follows: (a) the bare propagator $\langle\theta
\theta^{\prime}\rangle_0$ defined as
\begin{equation}
\langle\theta \theta^{\prime}\rangle_0=\langle\theta^{\prime}
\theta\rangle^*_0=\frac{1}{-i\omega+\nu_0 k^2},
\end{equation}
and b) the bare propagator for the velocity field $\langle v
v\rangle_0$ given directly by Eq.\,(\ref{corrvelo}), namely,
\begin{equation}
\langle v_i v_j\rangle_0 = P^{\rho}_{ij}({\bf k})
\tilde{D}^v(\omega, k),
\end{equation}
where $P^{\rho}_{ij}({\bf k})$ is the transverse projector defined
in the previous section by Eq.\,(\ref{projectorA}). Their graphical
representation is presented in Fig.\,\ref{propagators}.

The triple (interaction) vertex $-\theta^{\prime} v_j\partial_j
\theta = \theta^{\prime} v_j V_j \theta $ is present in
Fig.\,\ref{vertex}, where the momentum ${\bf k}$ is flowing into the
vertex via the auxiliary field $\theta^{\prime}$.

\section{UV renormalization and RG analysis}\label{sec:RG}

We start with the analysis of uv divergences, which is usually based
on the analysis of canonical dimensions (see, e.g.,
\cite{ZinnJustin,Vasiliev,Collins}). First of all, the dynamical
model (\ref{action3}), as well as all models of this type, is a
so-called two-scale model
\cite{Vasiliev,AdAnVa96,AdAnVa99,AdVaPi83}, i.e., the canonical
dimension of some quantity $F$ is described by two numbers, namely,
the momentum dimension $d^k_F$ and the frequency dimension
$d^{\omega}_F$. To find the dimensions of all quantities it is
appropriate to use the standard normalization conditions
$d^k_k=-d^k_x=1, d^{\omega}_{\omega}=-d^{\omega}_t=1,
d^{\omega}_k=d^{\omega}_x=d^k_{\omega}=d^k_t=0$, and the requirement
that each term of the action functional must be separately
dimensionless with respect to the momentum and frequency dimensions.
The total canonical dimension $d_F$ is then defined as $d_F=d^k_F+2
d^{\omega}_F$ [it is related to the fact that $\partial_t \propto
\nu_0 \partial^2$ in the free action (\ref{action3}) with choice of
zero canonical dimension for $\nu_0$]. In the framework of the
theory of renormalization  the total canonical dimension in
dynamical models plays the same role as the momentum dimension does
in static models.

The canonical dimensions of the model (\ref{action3}) cannot be
determined directly because it contains fewer terms than fields and
parameters. Thus one is faced with some kind of uncertainty in
calculation of canonical dimensions. This freedom is demonstrated by
the fact that parameter $h=|{\bf h}|$ can be eliminated from the
action (see Ref.\,\cite{Antonov99} for details). When $h$ is
eliminated from the action, which is equivalent to the assigning of
zero canonical dimension to it, the canonical dimensions of the
other quantities can be calculated unambiguously. They are present
in Table \ref{table1}, where also the canonical dimensions of the
renormalized parameters are shown.

The model is logarithmic at $\varepsilon=\eta=0$ (the coupling
constants $g_0$ and $u_0$ are dimensionless); therefore the uv
divergences in the correlation functions have the form of poles in
$\varepsilon, \eta$, and their linear combinations.

The quantity that plays a central role in the renormalization of the
model, namely, the role of the formal index of the uv divergence, is
the total canonical dimension of an arbitrary one-particle
irreducible correlation (Green's) function $\Gamma=\langle \Phi
\cdots \Phi \rangle_{1-ir}$. It is given as follows:
\begin{equation}
d_{\Gamma}=d^k_{\Gamma}+2 d^{\omega}_{\Gamma}=d+2-N_{\Phi} d_{\Phi},
\end{equation}
where $N_{\Phi}=\{N_{\theta},N_{\theta^{\prime}},N_{{\bf v}}\}$ are
the numbers of corresponding fields entering into the function
$\Gamma$, and summation over all types of fields is implied. It is
well known that superficial uv divergences, whose removal requires
counterterms, can be present only in those Green's functions
$\Gamma$ for which the total canonical index $d_{\Gamma}$ is a
non-negative integer.

A detailed analysis of divergences in the problem (\ref{action3})
was done in Ref.\,\cite{Antonov99} (see also
Refs.\,\cite{AdAnVa96,AdAnVa99}); therefore we shall present here
only basic facts and conclusions rather than to repeat all details.
First of all, every one-irreducible Green's function with
$N_{\theta^{\prime}}<N_{\theta}$ vanishes.
On the other hand, dimensional analysis based on Table I leads to
the conclusion that for any $d$, superficial divergences can be
present only in the one-irreducible Green's functions $\langle
\theta^{\prime} \theta \cdots \theta \rangle$ with only one field
$\theta^{\prime}$ ($N_{\theta^{\prime}}=1$) and an arbitrary number
$N_{\theta}$ of fields $\theta$. Therefore, in the model under
investigation, superficial divergences  can be found only in the
one-particle irreducible function $\langle \theta^{\prime} \theta
\rangle$. To remove them one needs to include into the action
functional a counterterm of the form $\theta^{\prime} \triangle
\theta$. Its inclusion is manifested by the multiplicative
renormalization of the bare parameters $g_0, u_0$, and $\nu_0$ in
the action functional (\ref{action3}):
\begin{equation}
\nu_0=\nu Z_{\nu},\,\,\, g_0=g \mu^{2\varepsilon+\eta}
Z_g,\,\,\,u_0=u\mu^{\eta} Z_u. \label{zetka}
\end{equation}
Here the dimensionless parameters $g,u$, and $\nu$ are the
renormalized counterparts of the corresponding bare ones, $\mu$ is
the renormalization mass (a scale setting parameter) in the minimal
subtraction (MS) scheme, and $Z_i=Z_i(g,u)$ are renormalization
constants.

\begin{table}
\caption{\label{table1} Canonical dimensions of the fields and
parameters of the model under consideration.}
\begin{ruledtabular}
\begin{tabular}{ccccccccc}
$F$ & ${\bf v}$ & $\theta$ & $\theta^{\prime}$ & $m, \Lambda, \mu$ &
$\nu_0, \nu$ & $g_0$ & $u_0$ & $g, u, h$ \\
\hline $d^k_F$ & -1 & -1 & $d+1$ & 1 & -2 & $2 \varepsilon +\eta$ &
$\eta$ & 0 \\
$d^{\omega}_F$ & 1 & 0 & 0 & 0 & 1 & 0 & 0 & 0 \\
$d_F$ & 1 & -1 & $d+1$ & 1 & 0 & $2 \varepsilon +\eta$ & $\eta$ & 0 \\
\end{tabular}
\end{ruledtabular}
\end{table}

The renormalized action functional has the following form:
\begin{eqnarray}
S^R(\Phi)&=&-\frac{1}{2} \int dt_1\,d^d{\bf x_1}\,dt_2\,d^d{\bf x_2}
\nonumber \\ && v_i(t_1,{\bf x_1}) [D_{ij}^v(t_1,{\bf
x_1};t_2,{\bf x_2})]^{-1} v_j(t_2,{\bf x_2}) \label{actionRen} \\
&+& \int dt\,d^d{\bf x}\,\, \theta^{\prime}\left[-\partial_t \theta
- v_i\partial_i\theta+\nu Z_1 \triangle\theta-{\bf h}\cdot{\bf v}
\right],\nonumber
\end{eqnarray}
where the correlator $D_{ij}^v$ is written in renormalized
parameters (in wave-number-frequency representation)
\begin{equation}
\tilde{D}_{ij}^v(\omega, k) = \frac{P^{\rho}_{ij}({\bf k}) g \nu^3
\mu^{2\varepsilon+\eta} k^{4-d-2\varepsilon-\eta}}{(i\omega+u \nu
\mu^{\eta} k^{2-\eta})(-i\omega+u \nu \mu^{\eta}
k^{2-\eta})}.\label{corrveloRen}
\end{equation}
By comparison of the renormalized action (\ref{actionRen}) with
definitions of the renormalization constants $Z_i$, $i=g,u,\nu$
[Eq.\,(\ref{zetka})] we come to the relations among  them:
\begin{equation}
Z_{\nu}=Z_1,\,\,\, Z_g=Z_{\nu}^{-3},\,\,\,
Z_u=Z_{\nu}^{-1}.\label{zetka1}
\end{equation}
The second and third relations are consequences of the absence of
the renormalization of the term with $D^v$ in the renormalized
action (\ref{actionRen}). Renormalization of the fields, the mass
parameter $m$, and the vector ${\bf h}$ is not needed, i.e.,
$Z_{\Phi}=1$ for all fields, $Z_m=1$,  and also $Z_{{\bf h}}=1$.

In what follows, we shall work with two-loop approximation to be
able to see the effects of helicity. The calculation of higher-order
corrections is more difficult in the models with turbulent velocity
field with finite correlation time than  in the cases with $\delta$
correlation in time. First of all, one has to calculate more
relevant Feynman diagrams in the same order of perturbation theory
(see below). A second and more problematic distinction is related to
the fact that the diagrams for the finite correlated case involve
two different dispersion laws, namely, $\omega \propto k^2$ for the
scalar field and $\omega \propto k^{2-\eta}$ for the velocity field.
This leads to complicated expressions for renormalization  constants
even in the simplest (one-loop) approximation
\cite{Antonov99,Antonov00}. But, as was discussed in
\cite{Antonov99,Antonov00,AdAnHo02}, this difficulty can be avoided
by the calculation of all renormalization constants in an arbitrary
specific choice of the exponents $\varepsilon$ and $\eta$ that
guarantees uv finiteness of the Feynman diagrams. From the point of
calculations the most suitable choice is to put $\eta=0$ and leave
$\varepsilon$ arbitrary.

Thus, the knowledge of the renormalization constants  for the
special choice $\eta=0$ is sufficient to obtain all important
quantities like the $\gamma$-functions, $\beta$-functions,
coordinates of fixed points, and critical dimensions.

This possibility is not automatic in general. In the model under
consideration it is the consequence of an analysis which shows that
in the MS scheme all the needed anomalous dimensions are independent
of the exponents $\varepsilon$ and $\eta$ in the two-loop
approximation. But in the three-loop approximation they can simply
appear \cite{AdAnHo02}.

In Ref.\,\cite{AdAnHo02} the two-loop corrections to the anomalous
exponents of model (\ref{action3}) without helicity were studied. We
shall continue those investigations including the effects of
helicity.

Now we can  continue with renormalization of the model. The relation
$S(\theta,\theta^{\prime},{\bf v},
e_0)=S^R(\theta,\theta^{\prime},{\bf v}, e, \mu)$, where $e_0$
stands for the complete set of bare parameters and $e$ stands for
the renormalized ones, leads to the relation $W(A, e_0)=W^R(A, e,
\mu)$ for the generating functional of connected Green's functions.
By application of the operator $\tilde{\cal{D}}_{\mu}\equiv\mu
\partial_{\mu}$ at fixed $e_0$ on both sides of the last equation
one obtains the basic RG differential equation
\begin{equation}
{\cal{D}}_{RG} W^R(A,e,\mu)=0, \label{RGE}
\end{equation}
where ${\cal{D}}_{RG}$ represents the operation
$\tilde{\cal{D}}_{\mu}$ written in the renormalized variables. Its
explicit form is
\begin{equation}
{\cal{D}}_{RG} = {\cal{D}}_{\mu} +
\beta_g(g,u)\partial_g+\beta_u(g,u)\partial_u-\gamma_{\nu}(g,u){\cal{D}}_{\nu},\label{RGoper}
\end{equation}
where we denote ${\cal{D}}_x\equiv x\partial_x$ for any variable $x$
and the RG functions (the $\beta$ and $\gamma$ functions) are given
by well-known definitions, and in our case, using  relations
(\ref{zetka1}) for renormalization constants, they have the
following form
\begin{eqnarray}
\gamma_{\nu}&\equiv& \tilde{\cal{D}}_{\mu} \ln Z_{\nu}, \label{gammanu}\\
\beta_g&\equiv&\tilde{\cal{D}}_{\mu} g =g
(-2\varepsilon-\eta+3\gamma_{\nu}), \label{betag}\\
\beta_u&\equiv&\tilde{\cal{D}}_{\mu} u =u
(-\eta+\gamma_{\nu}).\label{betau}
\end{eqnarray}

\begin{figure}[t]
     \vspace{0cm}
       \begin{center}
       \leavevmode
       \epsfxsize=8cm
       \epsffile{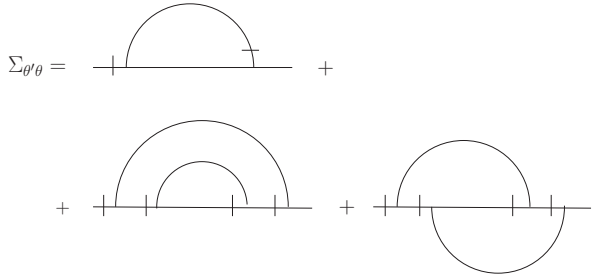}
   \end{center}
\vspace{-0.2cm} \caption{The one and two-loop diagrams that
contribute to the self-energy operator
$\Sigma_{\theta^{\prime}\theta}$. \label{fig3}}
\end{figure}

The renormalization constant $Z_{\nu}$ is determined by the
requirement that the one-irreducible Green's function $\langle
\theta^{\prime} \theta\rangle_{1-ir}$ must be uv finite when is
written in renormalized variables. In our case it means that they
have no singularities in the limit $\varepsilon, \eta\rightarrow0$.
The one-irreducible Green's function $\langle \theta^{\prime}
\theta\rangle_{1-ir}$ is related to the self-energy operator
$\Sigma_{\theta^{\prime}\theta}$ by the Dyson equation
\begin{equation}
\langle \theta^{\prime}\theta \rangle_{1-ir}=-i\omega+\nu_0 p^2 -
\Sigma_{\theta^{\prime}\theta}(\omega, p).\label{Dyson}
\end{equation}
Thus $Z_{\nu}$ is found from the requirement that the uv divergences
are canceled in Eq.\,(\ref{Dyson}) after the substitution $\nu_0=\nu
Z_{\nu}$. This determines $Z_{\nu}$ up to an uv-finite contribution,
which is fixed by the choice of the renormalization scheme. In the
MS scheme all the renormalization constants have the form (1 + poles
in $\varepsilon,\eta$ and their linear combinations). The
self-energy operator $\Sigma_{\theta^{\prime}\theta}$ is represented
by the corresponding one-irreducible diagrams. In contrast to the
rapid-change model, where only the one-loop diagram exists (it is
related to the fact that all higher-loop diagrams contain at least
one closed loop that is built up of only retarded propagators and
thus are automatically equal to zero), in the case with finite
correlations in time of the velocity field, higher-order corrections
are nonzero. In two-loop approximation the self-energy operator
$\Sigma_{\theta^{\prime}\theta}$ is defined by diagrams that are
shown in Fig.\,\ref{fig3}.

\input epsf

As was already mentioned, in our calculations we can put $\eta=0$.
This possibility essentially simplifies the evaluations of all
quantities \cite{Antonov99,Antonov00,AdAnHo02}. Then the singular
parts of the diagrams in Fig.\,\ref{fig3} have the following
analytical form (for calculational details see Appendix \ref{ApA}):
\begin{eqnarray}
A&=&-\frac{S_d}{(2\pi)^d}\frac{g \nu p^2}{4 u (1+u)}\frac{d-1}{d}
\left(\frac{\mu}{m}\right)^{2\varepsilon}\frac{1}{\varepsilon},\\
B_1&=&\frac{S_d^2}{(2\pi)^{2d}}\frac{g^2 \nu p^2}{16 u^2
(1+u)^3}\frac{(d-1)^2}{d^2}
\left(\frac{\mu}{m}\right)^{4\varepsilon} \nonumber
\\   \times && \hspace{-0.7cm}
\frac{1}{\varepsilon} \left( \frac{1}{2\varepsilon} +
\frac{{_2F_1}\left(1,1;2+\frac{d}{2};\frac{1}{(1+u)^2}\right)}{(d+2)(1+u)^2}
\right),
\end{eqnarray}
\begin{eqnarray}
B_2&=&\frac{S_d^2}{(2\pi)^{2d}}\frac{g^2 \nu p^2}{16 u^2
(1+u)^3}\frac{(d-1)}{d^2}\left(\frac{\mu}{m}\right)^{4\varepsilon}
\frac{1}{\varepsilon} \nonumber \\
&\times&\Bigg[\frac{{_2F_1}\left(1,1;2+\frac{d}{2};\frac{1}{(1+u)^2}\right)}{(d+2)(1+u)}
\label{b2}
\\
& &- \frac{(d-2)\pi\rho^2}{2}
\,{_2F_1}\left(\frac12,\frac12;1+\frac{d}{2};\frac{1}{(1+u)^2}\right)\Bigg],\nonumber
\end{eqnarray}
Expression $A$ is the result of the one-loop diagram, $B_1$ is the
result for the first two-loop diagram (the first diagram in the
second row in Fig.\,\ref{fig3}), and $B_2$ is the result for the
second two-loop graph (the second diagram in the second row in
Fig.\,\ref{fig3}). Here, $S_d=2 \pi^{d/2}/\Gamma(d/2)$ denotes the
$d$-dimensional sphere, ${_2F_1}(a,b,c,z)=1+\frac{a\,
b}{c\cdot1}z+\frac{a(a+1)b(b+1)}{c(c+1)\cdot1\cdot2}z^{2}+\ldots$
represents the corresponding hypergeometric function. In further
investigations the helical term with $\rho^2$ in $B_2$ has to be
taken with $d=3$ but for completeness we leave the $d$ dependence in
this part of $B_2$ in Eq.\,(\ref{b2}).

Finally, the renormalization constant $Z_{\nu}=Z_1$ is given as
follows:
\begin{eqnarray}
Z_{\nu}= 1&-& \frac{\bar{g}}{\varepsilon}\frac{d-1}{d}\frac{1}{4
u(1+u)} -
\frac{\bar{g}^2}{\varepsilon^2}\frac{(d-1)^2}{d^2}\frac{1}{32
u^2(1+u)^3} \nonumber \\
&+& \frac{\bar{g}^2}{\varepsilon} \frac{(d-1)(d+u)}{d^2(d+2)}
\frac{1}{16 u^2(1+u)^5} \nonumber \\
& &\times \, {_2F_1}\left(1,1;2+\frac{d}{2};\frac{1}{(1+u)^2}\right) \\
&-&\rho^2 \frac{\bar{g}^2}{\varepsilon} \frac{\pi}{144 u^2(1+u)^3}
\,{_2F_1}\left(\frac12,\frac12;\frac52;\frac{1}{(1+u)^2}\right),\nonumber
\end{eqnarray}
where in the helical part (the last line) we already substitute
$d=3$ and  denote $\bar{g}=g S_d/(2\pi)^d$.

Now using the definition of the anomalous dimension $\gamma_{\nu}$
in Eq.\,(\ref{gammanu}) one comes to the following expression:
\begin{equation}
\gamma_{\nu}=-2(\bar{g}{\cal A} + 2 \bar{g}^2 {\cal B}),
\label{gamanuu}
\end{equation}
where
\begin{equation}
{\cal A}= -\frac{d-1}{d}\frac{1}{4 u(1+u)}\label{AA}
\end{equation}
is the one-loop contribution to the anomalous dimension
$\gamma_{\nu}$ and the two-loop contribution is
\begin{eqnarray}
{\cal B} &=& \frac{(d-1)(d+u)}{16 d^2(d+2)u^2(1+u)^5}
\,{_2F_1}\left(1,1;2+\frac{d}{2};\frac{1}{(1+u)^2}\right)\nonumber
\\ & &- \frac{\pi \rho^2}{144 u^2(1+u)^3}
\,{_2F_1}\left(\frac12,\frac12;\frac52;\frac{1}{(1+u)^2}\right).\label{BB}
\end{eqnarray}

The issues of interest are especially the multiplicatively
renormalizable equal-time two-point quantities $G(r)$ (see, e.g.,
Ref.\,\cite{Antonov99}). Examples of such quantities are the
equal-time structure functions
\begin{equation}
S_{n}(r)\equiv\langle[\theta(t,{\bf x})-\theta(t,{\bf
x'})]^{n}\rangle \label{struc}
\end{equation}
in the inertial range, specified by the inequalities $l\sim
1/\Lambda \ll r \ll L=1/m$ ($l$ is an internal length). Here the
angular brackets $\langle \cdots \rangle$ mean the functional
average over fields $\Phi=\{\theta, \theta', {\bf v}\}$ with weight
$\exp S^R(\Phi).$ The infrared (ir) scaling behavior of the function
$G(r)$ (for $r/l\gg 1$ and any fixed $r/L$)
\begin{equation}
G(r)\simeq \nu_0^{d^{\omega}_G} l^{-d_G} (r/l)^{-\Delta_G} R(r/L)
\label{frscaling}
\end{equation}
is related to the existence of ir stable fixed points of the RG
equations (see the next section). In (\ref{frscaling})
$d^{\omega}_G$ and $d_G$ are corresponding canonical dimensions of
the function $G$, $R(r/L)$ is the so-called scaling function, which
cannot be determined by the RG equation (see, e.g.,
Ref.\,\cite{Vasiliev}), and $\Delta_G$ is the critical dimension
defined as
\begin{equation}
\Delta_G=d_G^k+\Delta_{\omega} d_G^{\omega} + \gamma_G^*.
\end{equation}
Here $\gamma_G^*$ is the fixed point value of the anomalous
dimension $\gamma_G\equiv \mu \partial_{\mu} \ln Z_G$, where $Z_G$
is the renormalization constant of the multiplicatively
renormalizable quantity $G$, i.e., $G=Z_G G^R$ \cite{Antonov00}, and
$\Delta_{\omega}=2-\gamma_{\nu}^*$ is the critical dimension of the
frequency with $\gamma_{\nu}$ which is defined in
Eq.\,(\ref{gamanuu}) (see also the next section).

On the other hand, the small-$r/L$ behavior of the scaling function
$R(r/L)$ can be studied using the Wilson OPE \cite{Vasiliev}. It
shows that, in the limit $r/L\to 0$, the function $R(r/L)$ can be
written in the following asymptotic form
\begin{equation}
R(r/L) = \sum_{F} C_{F}(r/L)\, (r/L)^{\Delta_F}, \label{ope}
\end{equation}
where $C_{F}$ are coefficients regular in $r/L$. In general,
summation is implied over certain  renormalized composite operators
$F$  with critical dimensions $\Delta_F$. In the case under
consideration the leading contribution is given by operators $F$
having the form $F_n= (\partial_i \theta \partial_i \theta)^n.$ In
Sec.\,\ref{sec:CompOper} we shall consider them in detail where the
complete two-loop calculation of the critical dimensions  of the
composite operators $F_n$ will be presented for arbitrary values of
$n$, $d$, $u$, and $\rho$.

\section{Fixed points and scaling regimes}\label{sec:ScalReg}

Possible scaling regimes of a renormalizable model are directly
given by the ir stable fixed points of the corresponding system of
RG equations \cite{ZinnJustin,Vasiliev}. The fixed point of the RG
equations is defined by $\beta$-functions, namely, by the
requirement of their vanishing. In our model, the coordinates $g_*,
u_*$ of the fixed points are found from the system of two equations
\begin{equation}
\beta_g(g_*,u_*)=\beta_u(g_*,u_*)=0.
\end{equation}
The beta functions $\beta_g$ and $\beta_u$ are defined in
Eqs.\,(\ref{betag}) and (\ref{betau}). To investigate the ir
stability of a fixed point it is enough to analyze the eigenvalues
of the matrix $\Omega$ of first derivatives:
\begin{equation}
\Omega_{ij}=\left(\begin{array}{cc}\partial \beta_g/\partial g &
\partial \beta_g/\partial u \\ \partial \beta_u/\partial g & \partial \beta_u/\partial u
\end{array}
\right).
\end{equation}
The ir asymptotic behavior is governed by the ir stable fixed
points, i.e., those for which both eigenvalues are positive.

\input epsf
   \begin{figure}[t]
     \vspace{-1.65cm}
       \begin{center}
       \leavevmode
       \epsfxsize=8cm
       \epsffile{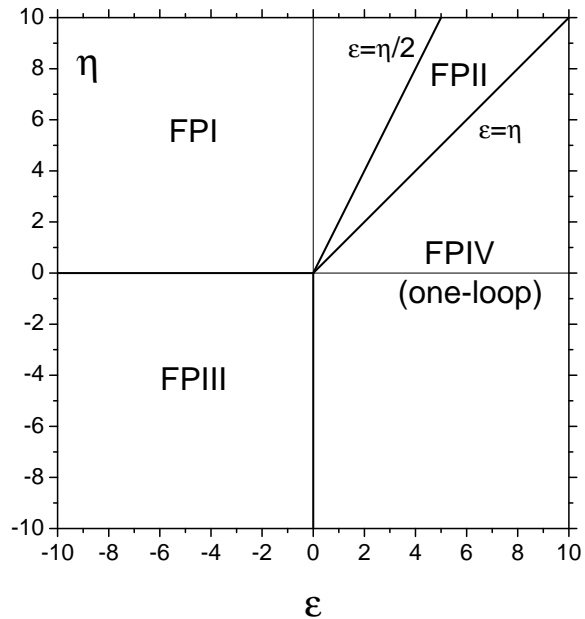}
   \end{center}
\vspace{-1.8cm} \caption{Regions of stability for the fixed points
in one-loop approximation. The regions of stability for fixed points
FPI, FPII, and FPIII are exact, i.e., not influenced by loop
corrections. The fixed point FPIV is shown in one-loop
approximation. The $d$ dependence of the FPIV in two-loop
approximation is shown in Fig.\,\ref{fig5} below.\label{fig4}}
\end{figure}

The possible scaling regimes of the model in one-loop approximation
were investigated in Ref.\,\cite{Antonov99}. Our first question is
how the two-loop approximation changes the picture of the "phase"
diagram of scaling regimes discussed in Ref.\,\cite{Antonov99}, and
the second one is what restrictions on this picture are given by
helicity (in the two-loop approximation). The two-loop approximation
in the model under our consideration without helicity was studied in
Ref.\,\cite{AdAnHo02} but the question of scaling regimes from the
two-loop approximation point of view was not discussed in detail.

\input epsf
   \begin{figure}[t]
     \vspace{-1.65cm}
       \begin{center}
       \leavevmode
       \epsfxsize=8cm
       \epsffile{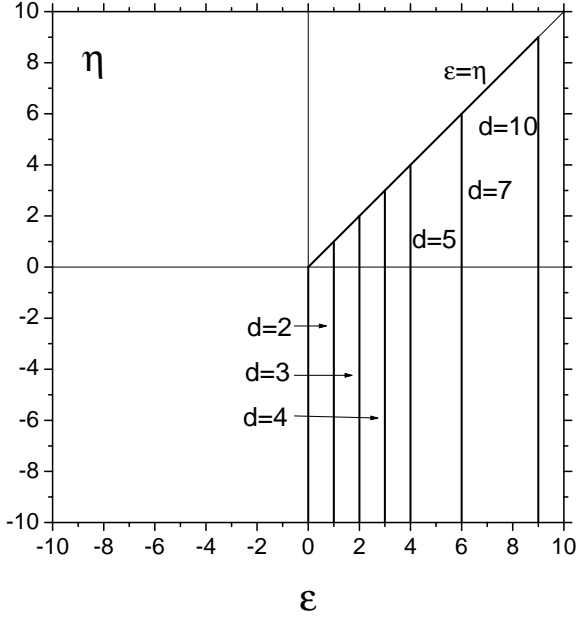}
   \end{center}
\vspace{-1.4cm} \caption{Regions of  stability for the fixed point
FPIV in the two-loop approximation without helicity for different
space dimensions $d$. The ir fixed point is stable in the region
given by the inequalities $\varepsilon>0, \varepsilon>\eta$, and
$\varepsilon<d-1$. \label{fig5}}
\end{figure}

First of all, we shall study the rapid-change limit
$u\rightarrow\infty$. In this regime, it is convenient to make a
transformation to new variables, namely, $w\equiv1/u$ and
$g^{\prime}\equiv g/u^2$, with the corresponding changes in the
$\beta$ functions:
\begin{eqnarray}
\beta_{g^{\prime}}&=&g^{\prime}(\eta-2\varepsilon +\gamma_{\nu}), \\
\beta_w&=&w(\eta-\gamma_{\nu}).
\end{eqnarray}
In this notation the anomalous dimension $\gamma_{\nu}$ obtains the
following form:
\begin{equation}
\gamma_{\nu}=-2(\bar{g}^{\prime}{\cal A^{\prime}} + 2
\bar{g}^{\prime 2} {\cal B^{\prime}}),
\end{equation}
where again $\bar{g}^{\prime}=g^{\prime} S_d/(2\pi)^d$. The one-loop
contribution ${\cal A^{\prime}}$ acquires the form
\begin{equation}
{\cal A^{\prime}}= -\frac{d-1}{d}\frac{1}{4(1+w)}
\end{equation}
and the two-loop correction ${\cal B^{\prime}}$ is
\begin{eqnarray}
{\cal B^{\prime}} &=& \frac{(d-1)(d w+1) w^2}{16 d^2(d+2)(1+w)^5}
\,{_2F_1}\left(1,1;2+\frac{d}{2};\frac{w^2}{(1+w)^2}\right)\nonumber
\\ & &- \frac{\pi \rho^2 w}{144 (1+w)^3}
\,{_2F_1}\left(\frac12,\frac12;\frac52;\frac{w^2}{(1+w)^2}\right).
\label{BsCiarou}
\end{eqnarray}
It is evident that in the rapid-change limit $w\rightarrow0$
($u\rightarrow\infty$) the two-loop contribution ${\cal B^{\prime}}$
is equal to zero. This is not surprising because in the rapid-change
model there are no higher-loop corrections to the self-energy
operator \cite{AdAnVa98+,AdAnBaKaVa01}; thus we are coming to the
one-loop result of Ref.\,\cite{Antonov99} with the anomalous
dimension $\gamma_{\nu}$ of the form
\begin{equation}
\gamma_{\nu}=\lim_{w\rightarrow 0}\frac{(d-1){\bar g^{\prime}}}{2
d(1+w)}=\frac{(d-1){\bar g^{\prime}}}{2 d}
\end{equation}
In this regime we have two fixed points denoted as FPI and FPII in
Ref.\,\cite{Antonov99}. The first fixed point is trivial, namely,
\begin{equation}
\mathrm{FPI}:\,\,\,\, w_*=g_*^{\prime}=0,
\end{equation}
with $\gamma_{\nu}^*=0$, and diagonal matrix $\Omega$ with
eigenvalues (diagonal elements)
\begin{equation}
\Omega_1=\eta,\,\,\,\,\,\Omega_2=\eta-2\varepsilon.
\end{equation}
The region of stability is shown in Fig.\,\ref{fig4}. The second
point is defined as
\begin{equation}
\mathrm{FPII}:\,\,\,\,w_*=0,\,\,\,
\bar{g}_*^{\prime}=\frac{2d}{d-1}(2\varepsilon-\eta),
\end{equation}
with $\gamma_{\nu}^*=2\varepsilon-\eta$. These are exact one-loop
expressions as a result of the nonexistence of the higher-loop
corrections [see discussion below (\ref{BsCiarou})]. That means that
they have no corrections of order $O(\varepsilon^2)$ and higher [we
work with the assumption that $\varepsilon \simeq \eta$; therefore
it also includes corrections of the type $O(\eta^2)$ and $O(\eta
\varepsilon)$]. The corresponding "stability matrix" is triangular
with diagonal elements (eigenvalues)
\begin{equation}
\Omega_1=2(\eta-\varepsilon),\,\,\,\,\Omega_2=2\varepsilon-\eta.
\end{equation}
The region of stability of this fixed point is shown in
Fig.\,\ref{fig4}.

Now let us analyze the "frozen regime" with frozen velocity field,
which is mathematically obtained from the model under consideration
in the limit $u \rightarrow 0$. To study this transition it is
appropriate to change the variable $g$  to the new variable
$g^{\prime\prime} \equiv g/u$ \cite{Antonov99}. Then the $\beta$
functions are transformed to the following:
\begin{eqnarray}
\beta_{g^{\prime\prime}}&=&g^{\prime\prime}(-2\varepsilon +2 \gamma_{\nu}), \label{betagu0}\\
\beta_u&=&u(-\eta+\gamma_{\nu}),\label{betauu0}
\end{eqnarray}
where the $\beta_u$ function is not changed, i.e., it is the same as
the initial one (\ref{betau}). In this notation the anomalous
dimension $\gamma_{\nu}$ has the form
\begin{equation}
\gamma_{\nu}=-2(\bar{g}^{\prime\prime}{\cal A^{\prime\prime}} + 2
\bar{g}^{\prime\prime 2} {\cal B^{\prime\prime}}),
\end{equation}
where, as obvious,  $\bar{g}^{\prime\prime}=g^{\prime\prime}
S_d/(2\pi)^d$. The one-loop part ${\cal A^{\prime\prime}}$ is now
defined as
\begin{equation}
{\cal A^{\prime\prime}}= -\frac{d-1}{d}\frac{1}{4(1+u)}
\end{equation}
and the two-loop one ${\cal B^{\prime\prime}}$ is given by
\begin{eqnarray}
{\cal B^{\prime\prime}} &=& \frac{(d-1)(d+u)}{16 d^2(d+2)(1+u)^5}
\,{_2F_1}\left(1,1;2+\frac{d}{2};\frac{1}{(1+u)^2}\right)\nonumber
\\ & &- \frac{\pi \rho^2}{144 (1+u)^3}
\,{_2F_1}\left(\frac12,\frac12;\frac52;\frac{1}{(1+u)^2}\right).
\label{Bs2Ciarou}
\end{eqnarray}
In the limit $u\rightarrow 0$ the functions ${\cal
A^{\prime\prime}}$ and ${\cal B^{\prime\prime}}$ obtain the
following forms
\begin{equation}
{\cal A^{\prime\prime}}_0= -\frac{d-1}{4 d}\label{As2Ciarouulim}
\end{equation}
and
\begin{equation}
{\cal B^{\prime\prime}}_0 =
\frac{(d-1){_2F_1}\left(1,1;2+\frac{d}{2};1\right)}{16 d (d+2)}  -
\frac{\pi \rho^2 {_2F_1}\left(\frac12,\frac12;\frac52;1\right)}{144}
. \label{Bs2Ciaroulim}
\end{equation}
The system of $\beta$ functions (\ref{betagu0}) and (\ref{betauu0})
exhibits two fixed points, denoted as FPIII and FPIV in
Ref.\,\cite{Antonov99}, related to the corresponding two scaling
regimes. One of them is trivial,
\begin{equation}
\mathrm{FPIII}:\,\,\,\, u_*=g_*^{\prime\prime}=0,
\end{equation}
with $\gamma_{\nu}^*=0$. The eigenvalues of the corresponding matrix
$\Omega$, which is diagonal in this case, are
\begin{equation}
\Omega_1=-2\varepsilon,\,\,\,\,\Omega_2=-\eta.
\end{equation}
Thus, this regime is ir stable only if both parameters $\varepsilon$
and $\eta$ are negative simultaneously as can be seen in
Fig.\,\ref{fig4}. The second, nontrivial, point is
\begin{equation}
\mathrm{FPIV}:\,\,\,\, u_*=0,\,\,\,\,
\bar{g}_*^{\prime\prime}=-\frac{\varepsilon}{2 {\cal
A^{\prime\prime}}_0}-\frac{{\cal B^{\prime\prime}}_0}{2 {\cal
A^{\prime\prime}}_0^2} \varepsilon^2,
\end{equation}
where ${\cal A^{\prime\prime}}_0$ and ${\cal B^{\prime\prime}}_0$
are defined in Eqs.\,(\ref{As2Ciarouulim}) and (\ref{Bs2Ciaroulim}),
respectively.

\input epsf
   \begin{figure}[t]
     \vspace{-1.65cm}
       \begin{center}
       \leavevmode
       \epsfxsize=8cm
       \epsffile{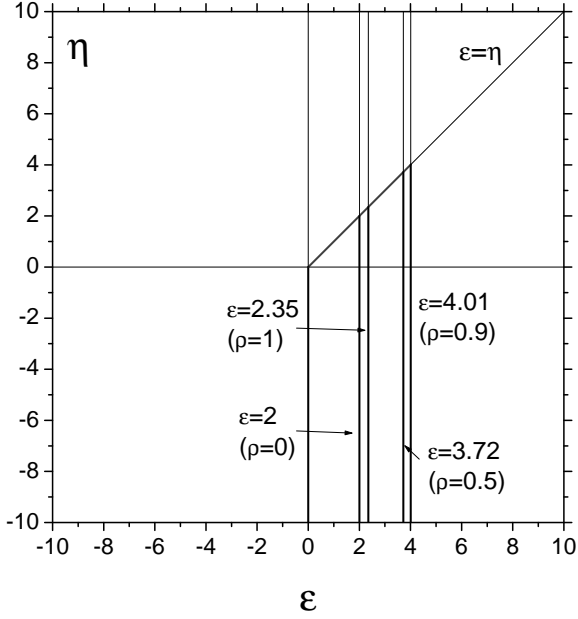}
   \end{center}
\vspace{-1.8cm} \caption{Regions of the stability for the fixed
point FPIV in two-loop approximation with helicity. The ir fixed
point is stable in the region given by the inequalities
$\varepsilon>0, \varepsilon>\eta$, and
$\varepsilon<\varepsilon_{\rho}$. \label{fig6}}
\end{figure}

\input epsf
   \begin{figure}[t]
     \vspace{-1.65cm}
       \begin{center}
       \leavevmode
       \epsfxsize=8cm
       \epsffile{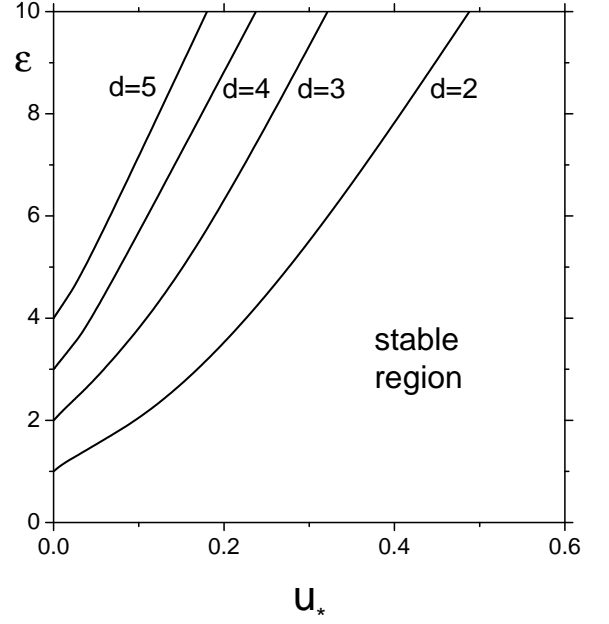}
   \end{center}
\vspace{-1.8cm} \caption{Regions of stability for the fixed point
FPV in the two-loop approximation without helicity. $d$ dependence
of the stability is shown. \label{fig7}}
\end{figure}

First, let us study the influence of the two-loop approximation on
this ir scaling regime without helicity in the general
$d$-dimensional case. We denote the corresponding fixed point as
FPIV$_{0}$, and its coordinates are
\begin{equation}
\mathrm{FPIV_0}:\,\,\,u_*=0,\,\,\,\,
\bar{g}_*^{\prime\prime}=\frac{2d}{d-1}\left(\varepsilon+\frac{1}{d-1}\varepsilon^2\right),
\end{equation}
with anomalous dimension $\gamma_{\nu}$ defined as
\begin{equation}
\gamma_{\nu}^*=\frac{d-1}{2d}\left(\bar{g}_*^{\prime\prime}-\frac{\bar{g}_*^{\prime\prime2}}{2d}\right)=\varepsilon,
\end{equation}
which is the exact one-loop result \cite{Antonov99}. The eigenvalues
of the matrix $\Omega$ (taken at the fixed point) are
\begin{equation}
\Omega_1=2\left(\varepsilon+\frac{1}{1-d}\varepsilon^2\right),\,\,\,
\Omega_2=\varepsilon-\eta.
\end{equation}
The eigenvalue $\Omega_2=\partial_u \beta_u|_*=-\eta+\gamma_{\nu}^*$
is also an exact one-loop result. The conditions
$\bar{g}_*^{\prime\prime}>0, \Omega_{1}>0$, and $\Omega_2>0$ for the
ir-stable fixed point lead to the following restrictions on the
values of the parameters $\varepsilon$ and $\eta$:
\begin{equation}
\varepsilon>0,\,\,\, \varepsilon>\eta,\,\,\,
\varepsilon<d-1.\label{cond1}
\end{equation}
The region of stability is shown in Fig.\,\ref{fig5}. The region of
stability of this ir fixed point increases when the dimension of the
coordinate space $d$ increases.

Now turn to the system with helicity. In this case the dimension of
the space is fixed for $d=3$. Thus, our starting conditions for a
stable ir fixed point of this type are obtained from the conditions
(\ref{cond1}) with the explicit value d=3: $\varepsilon>0,
\varepsilon>\eta, \varepsilon<2$. But they are valid only if
helicity is vanishing and could be changed when nonzero helicity is
present. Let us study this case. When helicity is present the fixed
point FPIV is given as
\begin{equation}
u_*=0,\,\,\,\,
\bar{g}_*^{\prime\prime}=3\varepsilon+\frac32\left(1-\frac{3\pi^2\rho^2}{16}\right)\varepsilon^2,\label{gpp}
\end{equation}
Therefore, in the helical case, the situation is a little bit more
complicated as a result of the competition between nonhelical and
helical terms within two-loop corrections. The matrix $\Omega$ is
triangular with diagonal elements (taken already at the fixed point)
\begin{eqnarray}
\Omega_1&=& 2\varepsilon+
\left(-1+\frac{3\pi^2\rho^2}{16}\right)\varepsilon^2,\label{lambda11}\\
\Omega_2&=&\varepsilon-\eta,
\end{eqnarray}
where explicit dependence of the eigenvalue $\Omega_1$ on the
parameter $\rho$ occures. The requirement to have positive values
for the parameter $\bar{g}_*^{\prime\prime}$, and at the same time
for the eigenvalues $\Omega_1, \Omega_2$ leads to the region of the
stable fixed point. The results are shown in Fig.\,\ref{fig6}. The
picture is rather complicated due to the very existence of the
critical absolute value of $\rho$
\begin{equation}
\rho_c=\frac{4}{\sqrt{3}\pi},
\end{equation}
which is defined from the condition of vanishing of the two-loop
corrections in Eqs.\,(\ref{gpp}) and (\ref{lambda11}):
\begin{equation}
\left(-1+\frac{3\pi^2\rho^2}{16}\right)=0.
\end{equation}
As was already discussed above, when helicity is not present, the
system exhibits this type of fixed point (and, of course, the
corresponding scaling behavior) in the region restricted by the
inequalities $\varepsilon>0, \varepsilon>\eta$, and $\varepsilon<2$.
The last condition changes when the helicity is switched on. The
important feature here is that the two-loop contributions to
$\bar{g}_*^{\prime\prime}$ and $\Omega_1$ have the same structure
but opposite sign. This leads to different sources of conditions in
the cases when $|\rho|<\rho_c$ and $|\rho|>\rho_c$, respectively. In
the situation with $|\rho|<\rho_c$ the positiveness of $\Omega_1$
plays a crucial role and one has the following region of stability
of the ir-fixed point FPIV:
\begin{equation}
\varepsilon>0,\,\,\, \varepsilon>\eta,\,\,\,
\varepsilon<\frac{32}{16-3\pi^2\rho^2}.\label{cond3}
\end{equation}
On the other hand, in the case with $|\rho|>\rho_c$, the principal
restriction on the ir-stable regime is yielded by the condition
$\bar{g}_*^{\prime\prime}>0$ with final ir-stable region defined as
\begin{equation}
\varepsilon>0,\,\,\, \varepsilon>\eta,\,\,\,
\varepsilon<\frac{32}{-16+3\pi^2\rho^2}.\label{cond4}
\end{equation}
Therefore, if we continuously increase the absolute value of the
helicity parameter $\rho$, the region of stability of the fixed
point defined by the last inequality in Eq.\,(\ref{cond3}) increases
too. This restriction vanishes completely when $|\rho|$ reaches the
critical value $\rho_c$, and the picture becomes the same as in the
one-loop approximation \cite{Antonov99}. In this rather specific
situation the two-loop influence on the region of stability of the
fixed point is exactly zero: the helical and nonhelical two-loop
contributions are canceled by each other. Then if the absolute value
of the parameter $\rho$ increases further, the last condition
appears again, namely, the third condition in Eq.\,(\ref{cond4}),
and the restriction becomes stronger when $|\rho|$ tends to its
maximal value, $|\rho|=1$. In this case of the maximal breaking of
mirror symmetry (maximal helicity), $|\rho|=1$, the region of the ir
stability of the fixed point is defined by the inequalities
$\varepsilon>0, \varepsilon=\eta$, and $\varepsilon<2.351$ (see
Fig.\,\ref{fig6}). It is interesting that the presence of helicity
in the system leads to the enlargement of the stability region.

\input epsf
   \begin{figure}[t]
     \vspace{-1.65cm}
       \begin{center}
       \leavevmode
       \epsfxsize=8cm
       \epsffile{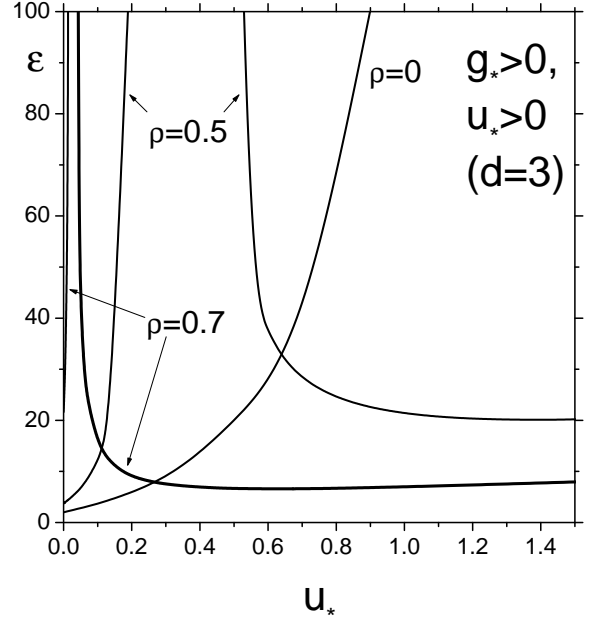}
   \end{center}
\vspace{-1.2cm} \caption{Regions of stability for the fixed point
FPV in the two-loop approximation with helicity in the situation
when $\rho<\rho_c=4/(3^{1/2}\pi)$.\label{Fig8}}
\end{figure}

Now let us turn to the most interesting scaling regime with a finite
value of the fixed point for the variable $u$. But by a short
analysis one immediately concludes that the system of equations (see
also \cite{Antonov99})
\begin{eqnarray}
\beta_g&=&g
(-2\varepsilon-\eta+3\gamma_{\nu})=0, \label{betag1}\\
\beta_u&=&u (-\eta+\gamma_{\nu})=0\label{betau1}
\end{eqnarray}
can be satisfied simultaneously for finite values of $g, u$ only in
the case when the parameter $\varepsilon$ is equal to $\eta$:
$\varepsilon=\eta$. In this case, the function $\beta_g$ is
proportional to the function $\beta_u$. As a result we have not one
fixed point of this type but a curve of fixed points in the $g-u$
plane. The value of the fixed point for the variable $g$  in the
two-loop approximation is given as follows (we denote it as in
Ref.\,\cite{Antonov99} as FPV):
\begin{equation}
\mathrm{FPV}:\,\,\,\bar{g}_*=-\frac{1}{2 {\cal
A}_*}\,\,\varepsilon-\frac{1}{2}\frac{{\cal B}_*}{{\cal
A}_*^3}\,\,\varepsilon^2,
\end{equation}
with the exact one-loop result for $\gamma_{\nu}^*=\varepsilon=\eta$
[this is already directly given by Eq.\,(\ref{betau1})]. Here ${\cal
A}_*$ and ${\cal B}_*$ are expressions ${\cal A}$ and ${\cal B}$
from Eqs.\,(\ref{AA}) and (\ref{BB}) which are taken at the fixed
point value $u_*$ of the variable $u$. The possible values of the
fixed point for the variable $u$
can be restricted (and will be restricted) as we shall discuss
below. The stability matrix $\Omega$ has the following eigenvalues:
\begin{equation}
\Omega_1=0,\,\,\, \Omega_2=3 \bar{g}^* \left(\frac{\partial
\gamma_{\nu}}{\partial g}\right)_*+u^*\left(\frac{\partial
\gamma_{\nu}}{\partial u}\right)_*.
\end{equation}
The vanishing of $\Omega_1$ is an exact result which is related to
the degeneracy of the system of Eqs.\,(\ref{betag1}) and
(\ref{betau1}) when nonzero solutions with respect to $g$ and $u$
are assumed, or, equivalently, it reflects the existence of a
marginal direction in the $g-u$ plane along the line of fixed
points.

\input epsf
   \begin{figure}[t]
     \vspace{-1.65cm}
       \begin{center}
       \leavevmode
       \epsfxsize=8cm
       \epsffile{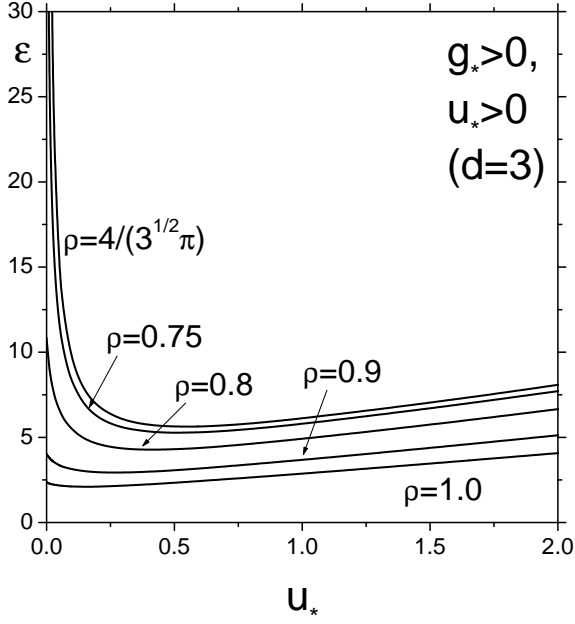}
   \end{center}
\vspace{-1.8cm} \caption{Regions of the stability for the fixed
point FPV in two-loop approximation with helicity in the situation
when $\rho \geq \rho_c=4/(3^{1/2}\pi)$. \label{Fig9}}
\end{figure}

We start the analysis of the last fixed point with the investigation
of the influence of the two-loop correction on the corresponding
scaling regime when helicity is not present in the system
($\rho=0$). In this situation it is interesting to determine the
dependence of the scaling regime on dimension $d$. The coordinates
of the possible fixed points are
\begin{eqnarray}
\bar{g}_*&=&\frac{2 d u_* (1+u_*)}{d-1}\varepsilon \nonumber
\\ &+&\frac{2 d u_*
(d+u_*){_2F_1}\left(1,1;2+\frac{d}{2};\frac{1}{(1+u_*)^2}\right)}{(d-1)^2(d+2)(1+u_*)^2}\varepsilon^2,
\end{eqnarray}
where $u_*$ is arbitrary for now. To have a positive value of the
fixed point for variables $g$ and $u$ one finds a restriction on the
parameter $\varepsilon$: $\varepsilon>0$. Possible restrictions on
the ir fixed point value of the variable $u$ can be found from the
condition $\Omega_2>0$. The explicit form of $\Omega_2$ is
\begin{eqnarray}
\Omega_2&=& \frac{2+u_*}{1+u_*}\varepsilon +\frac{\varepsilon^2}{(d-1)(d+2)(d+4)(1+u_*)^6}\nonumber \\
&\times& \bigg( (1+u_*)^2 (4 + d) (2 d (u_*-1) + (u_*-3)
u_*)\nonumber \\
& &\,\,\,\,\ \times \,\,
{_2F_1}\left(1,1;2+\frac{d}{2};\frac{1}{(1+u_*)^2}\right)
 \\
& &+ 4 u_* (d + u_*)
{_2F_1}\left(2,2;3+\frac{d}{2};\frac{1}{(1+u_*)^2}\right)\bigg).\nonumber
\end{eqnarray}
In Fig.\,\ref{fig7}, the regions of stability for the fixed point
FPV without helicity in the $\varepsilon-u$ plane for different
space dimensions $d$ are shown. It is interesting that in the
two-loop case a nontrivial $d$ dependence of ir stability appears,
in contrast to the one-loop approximation \cite{Antonov99}.

Now let us turn to the situation with helicity and investigate its
influence on the stability of the ir fixed point. In this case we
work in three-dimensional space; thus the coordinates of the fixed
point are defined by the following equation:
\begin{eqnarray}
\bar{g}_*&=& 3 u_* (1+u_*) \varepsilon + \frac{3 u_*
\varepsilon^2}{20
(1+u_*)^2} \nonumber \\
&\times&
\bigg(2(3+u_*){_2F_1}\left(1,1;\frac{7}{2};\frac{1}{(1+u_*)^2}\right)\label{fixh}
 \\
& & - 5\pi (1+u_*)^2 \rho^2
{_2F_1}\left(\frac12,\frac12;\frac{5}{2};\frac{1}{(1+u_*)^2}\right)\bigg).\nonumber
\end{eqnarray}
The competition between helical and nonhelical terms appears again,
which will lead to a nontrivial restriction for the fixed point
values of the  variable $u$ to have positive fixed values for
variable $g$. Next, the eigenvalue $\Omega_2$ of the matrix $\Omega$
is now
\begin{eqnarray}
\Omega_2&=& \frac{2+u_*}{1+u_*}\varepsilon+\frac{\varepsilon^2}{140
(1+u_*)^6} \nonumber \\
&\times& \bigg[ 8 u (3 + u)
{_2F_1}\left(2,2;\frac{9}{2};\frac{1}{(1+u_*)^2}\right) \nonumber
\\ & &\,\,\,\,+ 14 (1+u_*)^2 (u_* (3 + u_*)-6)\nonumber \\
& &\,\,\,\,\,\,\,\,\,\, \times \,
{_2F_1}\left(1,1;\frac{7}{2};\frac{1}{(1+u_*)^2}\right) \nonumber \\
& &\,\,\,\,\,+  7\pi \rho^2(1+u_*)^2 \nonumber \\
& &\,\,\,\,\,\,\times \bigg(10 (1+u_*)^2
\left(\frac12,\frac12;\frac{5}{2};\frac{1}{(1+u_*)^2}\right)
\nonumber \\
& &\,\,\,\,\,\,\,\,\,\,\,\, - u_*
\left(\frac32,\frac32;\frac{7}{2};\frac{1}{(1+u_*)^2}\right)\bigg)
\bigg]\label{lambda2k}
\end{eqnarray}
with a nontrivial helical part that plays an important role in
determination of the region of ir stability of the fixed point.

It cannot be seen immediately from Eqs.\,(\ref{fixh}) and
(\ref{lambda2k}) but numerical analysis shows that again an
important role is played by $\rho_c=4/(\sqrt{3}\pi)$. First let us
study the case when $|\rho|<\rho_c$. The corresponding region of
stable ir fixed points with $g_*>0$ is shown in Fig.\,\ref{Fig8}. In
the case when helicity is not present ($\rho=0$, see the
corresponding curve in Fig.\,\ref{Fig8}), the only restriction is
given by the condition that $\Omega_2>0$; on the other hand, the
condition $g_*>0$ is satisfied without restriction on the parameter
space. When arbitrarily small helicity is present, i.e., $\rho>0$, a
restriction related to positiveness of $g_*$ arises and is stronger
when $|\rho|$ is increasing (the right curve for the concrete value
of $\rho$ in Fig.\,\ref{Fig8}) and comes to play the dominant role.
At the same time, with increasing of $|\rho|$ the importance of the
positiveness of the eigenvalue $\Omega_2$ decreases (the left curve
for the concrete value of $\rho$ in Fig.\,\ref{Fig8}). For a given
$|\rho|<\rho_c$ there exists an interval of values of the variable
$u_*$ for which there is no restriction on the value of the
parameter $\varepsilon$. For example, for $|\rho|=0.1$, it is
$1.128<u_*<13.502$, for $|\rho|=0.5$, $0.217<u_*<0.394$, and for
$|\rho|=0.7$, $0.019<u_*<0.029$. Now turn to the case
$|\rho|\geq\rho_c$. When $|\rho|$ acquires its critical value
$\rho_c$, the ir fixed point is stable for all values of $u_*>0$ and
$\varepsilon>0$, i.e., the condition $\Omega_2>0$ becomes satisfied
without restrictions on the parameter space. On the other hand, the
condition $g_*>0$ yields a strong enough restriction and it becomes
stronger when $|\rho|$ tends to its maximal value $|\rho|=1$ as it
can be seen in Fig.\,\ref{Fig9}).

The most important conclusion of our two-loop investigation of the
model is the fact that the possible restrictions on the regions of
stability of ir fixed points are "pressed" to the region with rather
large values of $\epsilon$, namely, $\varepsilon\geq 2$, and do not
disturb the regions with relatively small $\varepsilon$. For
example, the Kolmogorov point ($\varepsilon=\eta=4/3$) is not
influenced.

As was already discussed (see the previous section) if $F$ denotes
some multiplicatively renormalized quantity (a parameter, a field,
or a composite operator) then its critical dimension is given by the
expression
\begin{equation}
\Delta[F]\equiv\Delta_F=d_F^k+\Delta_{\omega}d_F^{\omega}+\gamma^*_F\,,\label{deltaa}
\end{equation}
see, e.g., Refs.\,\cite{AdAnVa96,AdAnVa99,Vasiliev} for details. In
Eq.\,(\ref{deltaa}) $d_F^{k}$ and $d_F^{\omega}$ are the canonical
dimensions of $F$, $\Delta_{\omega}=2-\gamma^*_{\nu}$ is the
critical dimension of frequency, and $\gamma^*_F$ is the value of
the anomalous dimension $\gamma_F\equiv\tilde{\cal D}_{\mu}\ln Z_F$
at the corresponding fixed point. Because the anomalous dimension
$\gamma_{\nu}$ is already exact  for all fixed points at one-loop
level, the critical dimensions of frequency $\omega$ and of fields
$\Phi\equiv\{{\bf v}, \theta, \theta^{\prime}\}$  are also found
exactly at one-loop level approximation \cite{Antonov99}. In our
notation they read
\begin{eqnarray}
\Delta_{\omega}&=&2-2\varepsilon+\eta\,\,\,\,\,\,\,\,\,\,\,\mathrm{for}\,\,\mathrm{FPII},\nonumber \\
\Delta_{\omega}&=&2-\varepsilon\,\,\,\,\,\,\,\,\,\,\,\,\,\,\,\,\,\,\,\,\,\,\,\,\,\mathrm{for}\,\,\mathrm{FPIV}, \\
\Delta_{\omega}&=&2-\varepsilon=2-\eta\,\,\,\,\mathrm{for}\,\,\mathrm{FPV},\nonumber
\end{eqnarray}
and
\begin{equation}
\Delta_{{\bf
v}}=1-\gamma^*_{\nu},\,\,\,\Delta_{\theta}=-1,\,\,\,\Delta_{\theta^{\prime}}=d+1.
\end{equation}

Now let us consider some equal-time two-point quantity $F(r)$ that
depends on a single distance parameter $r$ which is multiplicatively
renormalizable ($F=Z_F F^R$, where $Z_F$ is the corresponding
renormalization constant). Then the renormalized function $F^R$ must
satisfy the RG equation of the form
\begin{equation}
({\cal D}_{RG}+\gamma_F)F(r)=0,
\end{equation}
with operator ${\cal D}_{RG}$ given explicitly in
Eq.\,(\ref{RGoper}) and usually  $\gamma_F\equiv {\tilde {\cal
D}}_{\mu} \ln Z_F$. The difference between the functions $F$ and
$F^R$ is only in the normalization, choice of parameters (bare or
renormalized), and related to this choice the form of the
perturbation theory (in $g_0$ or in $g$). The existence of a
nontrivial ir stable fixed point means that in the ir asymptotic
region $ r/l \gg 1$ and any fixed $r/L$ the function $F(r)$ takes on
the self-similar form
\begin{equation}
F(r)\simeq\nu_0^{d_F^{\omega}} l^{-d_F}(r/l)^{-\Delta_F}
f(r/L),\label{Fr}
\end{equation}
where the values of the critical dimensions correspond to the given
fixed point (see above in this section and Table\,\ref{table1}), and
$f$ is some scaling function whose explicit form is not determined
by the RG equation itself. The dependence of the scaling functions
on the argument $r/L$ in the region $r/L \ll 1$ can be studied using
the well-known Wilson operator product expansion (also known as the
short-distance expansion)
\cite{ZinnJustin,Vasiliev,AdAnVa96,AdAnVa99}. The OPE analysis will
be studied in Sec.\,\ref{sec:CompOper}.

\section{\label{sec:EffDiff}Effective diffusivity}

One of the interesting objects from the theoretical as well as
experimental point of view is the so-called effective diffusivity
$\bar \nu$. In this section let us briefly investigate the effective
diffusivity $\bar \nu$, which replaces the initial molecular
diffusivity $\nu_0$ in Eq.\,(\ref{scalar1}) due to the interaction
of a scalar field $\theta$ with the random velocity field ${\bf v}$.
The molecular diffusivity $\nu_0$ governs exponential damping in
time of all fluctuations in the system in the lowest approximation,
which is given by the propagator (response function)
\begin{eqnarray}
G(t-t',{\bf k})&=&\langle \theta(t,
{\bf{k}})\theta^{\prime}(t^{\prime}, {\bf{k}}) \rangle_0 \nonumber
\\ &=& \theta(t-t^{\prime})\exp(-\nu_0 k^2 (t-t^{\prime})).
\end{eqnarray}
 Analogously,
the effective diffusivity $\bar \nu$ governs exponential damping of
all fluctuations described by the full response function, which is
defined by the Dyson equation (\ref{Dyson}). Its explicit expression
can be obtained by the RG approach. In accordance with general rules
of the RG (see, e.g., Ref.\,\cite{Vasiliev}) all principal
parameters of the model $g_0,u_0$, and $\nu_0$ are replaced by their
effective (running) counterparts, which satisfy the Gell-Mann-Low RG
equations
\begin{equation}
s \frac{d \bar g}{d s}=\beta_{g}(\bar g,\bar u)\,,\, s \frac{d \bar
u}{d s}=\beta_{u}(\bar g, \bar u),\ \label{gu}
\end{equation}
\begin{equation}
s \frac{d \bar \nu}{d s}= -\bar \nu \gamma_{\nu}(\bar g,\bar u)\, ,
\label{nu}
\end{equation}
with initial conditions $\bar g|_{s=1}=g, \bar u|_{s=1}=u, \bar
\nu|_{s=1}=\nu$. Here $s=k/\mu$,  $\beta$, and $\gamma$ functions
are defined in Eqs.\,(\ref{gammanu}) - (\ref{betau}) and  all
running parameters clearly depend on the variable $s$.
Straightforward integration (at least numerical) of Eqs.\,(\ref{gu})
gives a method to find their fixed points. Instead, one very often
solves the set of equations $\beta_{g}( g_*, u_*)=\beta_{u}( g_*,
u_*)=0$ which defines all fixed points $g_*, u_*$. This last
approach was used above when we classified all fixed points. Due to
the special form of the $\beta$ functions (\ref{betag}),
(\ref{betau}) we are able to solve Eq.\,(\ref{nu}) analytically.
Using Eqs.\,(\ref{gu}) and (\ref{betag}) one immediately rewrites
(\ref{nu}) in the form
\begin{equation}
s \frac{d\bar  \nu}{\bar \nu}= \frac{\gamma_{\nu}}{2\varepsilon+\eta
- 3 \gamma_{\nu}}\frac{d \bar g}{\bar g}, \label{nu1}
\end{equation}
which can be easily integrated. Using initial conditions the
solution acquires the form
\begin{equation}
\bar \nu = \left(\frac{g\nu^3}{\bar g s^{2\epsilon +
\eta}}\right)^{1/3}= \left(\frac{D_0}{\bar g k^{2\epsilon +
\eta}}\right)^{1/3}\, , \label{nue}
\end{equation}
where to obtain the last expression we used the equations
$g\mu^{2\epsilon+\eta} \nu^3=g_0\nu_0^3 =D_0$. We emphasize that the
above solution is exact, i.e., the exponent $2\epsilon+\eta$ is
exact too. However, in the infrared region $k<< \Lambda \sim
l^{-1}$, $\bar g \rightarrow g_*,$ which can be calculated only
pertubatively. In the two-loop approximation $g_*=
g^{(1)}_*\varepsilon + g^{(2)}_*\varepsilon^2$ and after the Taylor
expansion of $g_*^{1/3}$ in Eq.\,(\ref{nue}) we obtain
\begin{equation}
\bar \nu \approx \nu_*
\left(\frac{D_0}{g_*^{(1)}\varepsilon}\right)^{1/3}
k^{-\frac{2\epsilon + \eta}{3}}\,,\qquad \nu_*\equiv
1-\frac{g_*^{(2)}\varepsilon}{3 g_*^{(1)}}\,. \label{nue1}
\end{equation}
Recall that for Kolmogorov values $\varepsilon =\eta=4/3$ the
exponent in (\ref{nue1}) becomes equal to $-4/3.$ Let us estimate
the contribution of helicity to the effective diffusivity in the
nontrivial point above denoted as FPV (\ref{fixh}). At this point
$\varepsilon=\eta$ $[(2\varepsilon+\eta)/3=\varepsilon]$ and
\begin{eqnarray}
\nu_*&=&1- \frac{\varepsilon} {12(1+u_*)} \nonumber \\
&&\times
\Biggl(\frac{2(3+u_*)}{5(1+u_*)^2}{_2F_1}\left(1,1;\frac{7}{2};\frac{1}{(1+u_*)^2}\right)
 \nonumber \\ && \hspace{0.7cm} - \pi \rho^2
{_2F_1}\left(\frac12,\frac12;\frac{5}{2};\frac{1}{(1+u_*)^2}\right)
\Biggr).\label{nukon}
\end{eqnarray}

\input epsf
   \begin{figure}[t]
     \vspace{-1.65cm}
       \begin{center}
       \leavevmode
       \epsfxsize=8cm
       \epsffile{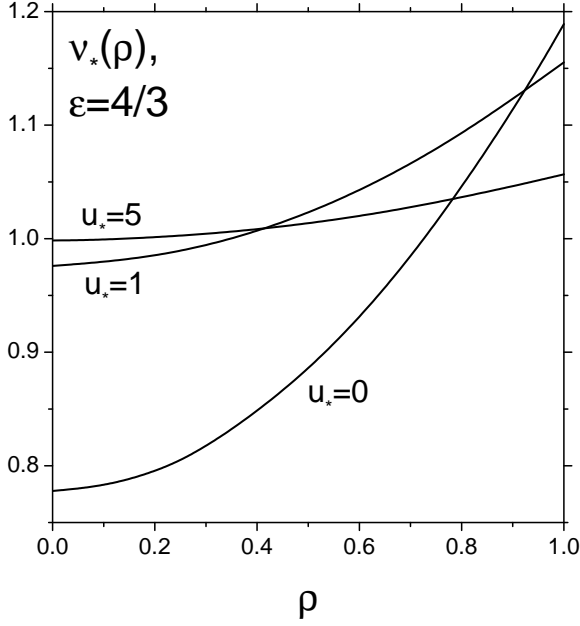}
   \end{center}
\vspace{-1.8cm} \caption{The dependence of $\nu_*$ on the helicity
parameter $\rho$ for definite ir fixed point values $u_*$ of the
parameter $u$. \label{fig10}}
\end{figure}

\input epsf
   \begin{figure}[t]
     \vspace{-1.65cm}
       \begin{center}
       \leavevmode
       \epsfxsize=8cm
       \epsffile{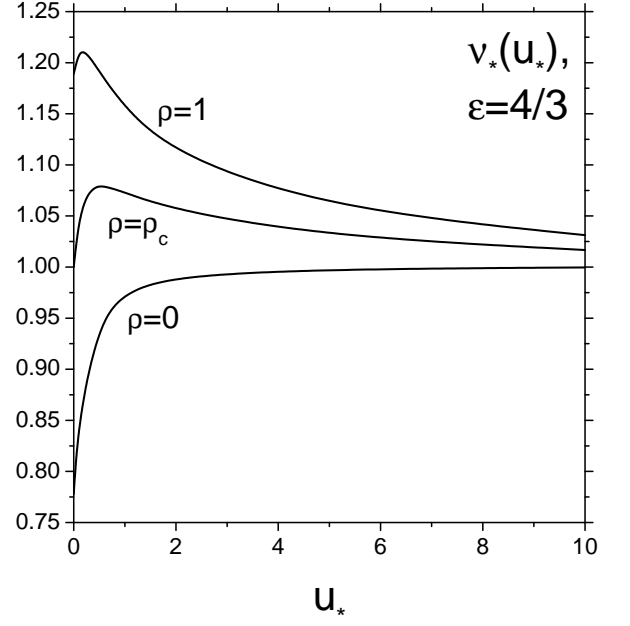}
   \end{center}
\vspace{-1.8cm} \caption{The dependence of $\nu_*$ on the ir fixed
point $u_*$ for concrete values of the helicity parameter $\rho$.
\label{fig11}}
\end{figure}

In Figs.\,\ref{fig10} and \ref{fig11} the dependence of the $\nu_*$
on the helicity parameter $\rho$ and the ir fixed point $u_*$ of the
parameter $u$ is shown. As one can see from these figures when
$u_*\rightarrow\infty$ (the rapid-change model limit) the two-loop
corrections to $\nu_*=1$ vanish. Such behavior is related to the
fact, which was already stressed in the paper, that within the
rapid-change model there are no two- and higher-loop corrections at
all. On the other hand, the largest two-loop corrections to $\nu_*$
are given in the frozen-velocity-field limit ($u_* \rightarrow 0$).
It is interesting that for all finite values of the parameter $u_*$
there exists a value of the helicity parameter $\rho$ for which the
two-loop contributions to $\nu_*$ are canceled. For example, for the
frozen-velocity-field limit ($u_*=0$) this situation arises when the
helicity parameter $\rho$ is equal to its critical value
$\rho_c=4/(\sqrt{3}\pi)$ (this situation can be seen in
Fig.\,\ref{fig11}). It is again the result of the competition
between the nonhelical and helical parts of the the two-loop
corrections as is shown in Eq.\,(\ref{nukon}). A further important
feature of the expression (\ref{nukon}) is that it is linear in the
parameter $\varepsilon$. Thus, when one varies the value of
$\varepsilon$ the picture is the same as in Figs.\,\ref{fig10} and
\ref{fig11} and only the scale of corrections is changed. In
Figs.\,\ref{fig10} and \ref{fig11} we have shown the situation for
the most interesting case when $\varepsilon$ is equal to its
Kolmogorov value, namely, $\varepsilon=4/3$.

\section{Operator-product expansion, Critical
dimensions of composite operators, and Anomalous
scaling}\label{sec:CompOper}

\subsection{Operator-product expansion}

Let us now study the behavior of the scaling function in
Eq.\,(\ref{Fr}). According to the OPE
\cite{ZinnJustin,Vasiliev,AdAnVa96,AdAnVa99}, the equal-time product
$F_1(x^{\prime})F_2(x^{\prime\prime})$ of two renormalized composite
operators \footnote{By definition we use the term "composite
operator" for any local monomial or polynomial constructed from
primary fields and their derivatives at a single point $x\equiv(t,
{\bf x})$. The constructions $\theta^n(x)$ and $[\partial_i
\theta(x)
\partial_i \theta(x)]^n$ are typical examples.} at ${\bf x}=({\bf x^{\prime}}+{\bf
x^{\prime\prime}})/2=const$ and ${\bf r}={\bf x^{\prime}}-{\bf
x^{\prime\prime}} \rightarrow 0$ can be written in the following
form:
\begin{equation}
F_1(x^{\prime})F_2(x^{\prime\prime})=\sum_i C_{F_i}({\bf r})
F_i({\bf x},t), \label{fff}
\end{equation}
where the summation is taken over all possible renormalized local
composite operators $F_i$ allowed by symmetry with definite critical
dimensions $\Delta_{F_i}$, and the functions $C_{F_i}$ are the
corresponding Wilson coefficients regular in $L^{-2}$. The
renormalized correlation function $\langle
F_1(x^{\prime})F_2(x^{\prime\prime}) \rangle$ can now be found by
averaging Eq.\,(\ref{fff}) with the weight $\exp S^R$ with $S^R$
from Eq.\,(\ref{actionRen}). The quantities $\langle F_i \rangle$
appear on the right-hand side, and their asymptotic behavior in the
limit $L^{-1} \rightarrow 0$ is then found from the corresponding RG
equations and has the form $\langle F_i \rangle \varpropto
L^{-\Delta_{F_i}}$.

From the OPE (\ref{fff}) one can find that the scaling function
$f(r/L)$ in the representation (\ref{Fr}) for the correlation
function $F_1(x^{\prime})F_2(x^{\prime\prime})$ has the form
\begin{equation}
f(r/L)=\sum_{i} A_{F_i} (r/L)^{\Delta_{F_i}},\label{ff1}
\end{equation}
where the coefficients $A_{F_i}$ are regular in $(r/L)^2$.

The principal feature of the turbulence models is the existence of
operators with negative critical dimensions (also known as
"dangerous" operators)
\cite{AdAnVa98+,AdAn98,AdAnVa96,AdAnVa99,Vasiliev}. Their presence
in the OPE determines the ir behavior of the scaling functions and
leads to their singular dependence on $L$ when $r/L \rightarrow0$.
At this point the turbulence models are crucially different from the
models of critical phenomena, where the leading contribution to the
representation (\ref{ff1}) is given by the simplest operator $F=1$
with the dimension $\Delta_F=0$, and the other operators determine
only the corrections that vanish for $r/L \rightarrow0$.

If the spectrum of the dimensions $\Delta_{F_i}$ for a given scaling
function is bounded from below, the leading term of its behavior for
$r/L \rightarrow 0$ is given by the minimal dimension. As was
discussed in Ref.\,\cite{Antonov99}, the model under consideration
belongs to this case for small enough values of the exponents
$\varepsilon, \eta$.

In what follows, we shall concentrate on the equal-time structure
functions of the scalar field defined as
\begin{equation}
S_n(r)\equiv\langle[\theta({\bf x},t)-\theta({\bf
x^{\prime}})]^n\rangle, r=|{\bf x}-{\bf x^{\prime}}|,\label{Sn}
\end{equation}
which are also interesting from the experimental point of view. The
representation (\ref{Fr}) is valid with the dimensions
$d_F^{\omega}=0$ and $d_F=\Delta_F=n\Delta_{\theta}=-n$. In general,
not only do the operators that are present in the corresponding
Taylor expansion enter into the OPE but also all possible operators
that admix with them in renormalization. In the present model the
leading contribution of the Taylor expansion for the structure
functions (\ref{Sn}) is given by the tensor composite operators
constructed solely of the scalar gradients
\begin{equation}
F[n,p]\equiv \partial_{i_1}\theta \cdots \partial_{i_p}\theta
(\partial_{i}\theta
\partial_{i} \theta)^l, \label{composite}
\end{equation}
where $n=p+2l$ is the total number of fields $\theta$ entering the
operator and $p$ is the number of free vector indices.

\subsection{ Composite operators $F[n,p]$:
renormalization and critical dimensions}

As the composite operators (\ref{composite}) play a central role in
what follows, let us briefly discuss their renormalization. A
complete and detailed discussion of the renormalization of the
composite operators is given in Ref.\,\cite{AdAnBaKaVa01}.
Therefore, we shall show only the basic moments necessary to present
explicit expressions for composite operators.

The necessity of additional renormalization of the composite
operators (\ref{composite}) is related to the fact that the
coincidence of the field arguments in Green's functions containing
them leads to additional uv divergences. These divergences must be
removed by special kinds of renormalization procedures which can be
found, e.g., in \cite{ZinnJustin,Vasiliev,Collins}, where their
renormalization is studied in general. The renormalization of
composite operators in models of turbulence is discussed in
Refs.\,\cite{AdVaPi83,AdAnVa99}. Typically, the composite operators
are mixed under renormalization. This means that renormalized
operators (which are uv finite) are linear combinations of
unrenormalized ones. In our case, the most important fact in
renormalization of composite operators $F[n,p]$ is that they mix
only with each other during the renormalization procedure; therefore
the corresponding matrix of renormalization constants $Z$ is found
from the condition of multiplicative renormalization which can be
written as follows:
\begin{equation}
F[n,p]=Z_{[n,p][n^{\prime},p^{\prime}]}F_R[n^{\prime},p^{\prime}]\,,\label{renFk}
\end{equation}
where $F_R$ denotes the renormalized counterpart of the composite
operator $F$. It is standard to define the matrix of corresponding
anomalous dimensions as
\begin{equation}
\gamma_{[n,p][n^{\prime},p^{\prime}]}=Z^{-1}_{[n,p][n^{\prime\prime},p^{\prime\prime}]}
{\tilde{\cal{D}}}_{\mu}
Z_{[n^{\prime\prime},p^{\prime\prime}][n^{\prime},p^{\prime}]}.\label{anomdimF}
\end{equation}

After the corresponding analysis of diagrams (for details see, e.g.,
Ref.\,\cite{Antonov99}) it can be shown that the renormalization
matrix $Z_{[n,p][n^{\prime},p^{\prime}]}$ in Eq.\,(\ref{renFk}) is
triangular; therefore, the matrix of anomalous dimensions
(\ref{anomdimF}) is also triangular. Thus, the anomalous dimensions
[the eigenvalues of the matrix (\ref{anomdimF})] is directly
determined by the diagonal elements of the matrix (\ref{renFk}),
namely,
\begin{equation}
\gamma[n,p]={\tilde{\cal{D}}}_{\mu} Z_{[n,p][n,p]}.\label{anomdimF1}
\end{equation}

\begin{figure}[t]
     \vspace{0cm}
       \begin{center}
       \leavevmode
       \epsfxsize=8cm
       \epsffile{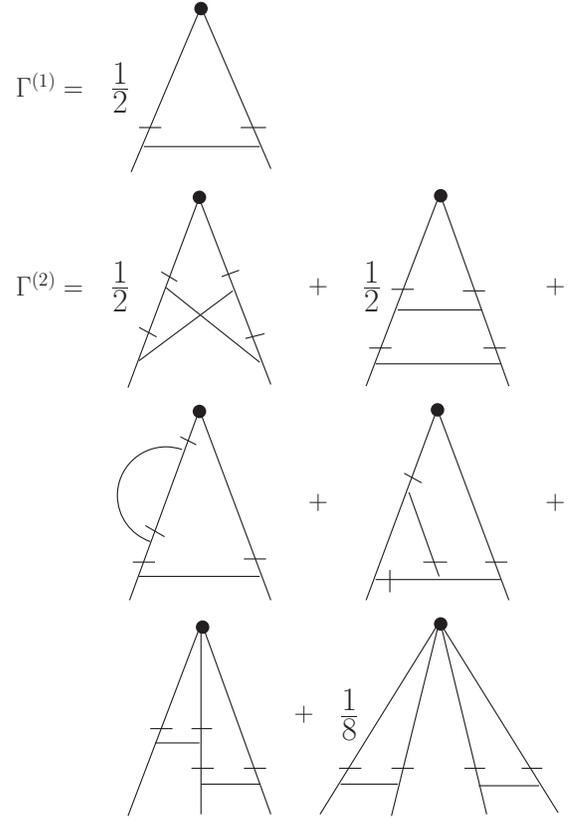}
   \end{center}
\vspace{0cm} \caption{The Feynman diagrams for the function
$\Gamma_{n,p}$ in two-loop approximation. The Feynman rules are the
same as in Sec.\,\ref{sec:Field}. The black circle is defined in the
text of the present section. The one-loop diagram we denote as
$K_1$, the two-loop diagrams we denote as $K_{2a}$  (the first
diagram), $K_{2b}$  (the second diagram), $K_{2c}$  (the first
diagram in the second row), $K_{2d}$  (the second diagram in the
second row), $K_{2e}$  (the first diagram in the third row), and
$K_{2f}$  (the second diagram in the third row). \label{fig12}}
\end{figure}

Our following aim is the calculation of the diagonal elements
$Z_{[n,p][n,p]}$ of the renormalization constants matrix
$Z_{[n,p][n^{\prime},p^{\prime}]}$. If we denote the generating
functional of the one-irreducible Green's functions with one
composite operator $F[n,p]$ [given in Eq.\,(\ref{composite})] and
any number of fields $\theta$ as $\Gamma(x;\theta)$ then we are
interested in part of it, namely, the $\theta^n$ term of the
expansion of $\Gamma(x;\theta)$ in $\theta$,  which will be denoted
as $\Gamma_{n,p}(x;\theta)$. Its analytical form is the following:
\begin{eqnarray}
\Gamma_{n,p}(x;\theta)&=&\frac{1}{n!}\int dx_{1}\cdots\int
dx_{n}\,\theta(x_{1})\cdots\theta(x_{n}) \nonumber \\ && \times
\langle
F[n,p](x)\theta(x_{1})\cdots\theta(x_{n})\rangle_{\textrm{1-ir}}
\nonumber \\ &\equiv& \frac{1}{n!}\int dx_{1}\cdots\int
dx_{n}\,\theta(x_{1})\cdots\theta(x_{n}) \nonumber \\
&& \times \Gamma_{n,p}(x;x_1,\ldots,x_n),\label{Gamma1}
\end{eqnarray}
where $\theta(x)$ is the functional argument, the "classical
counterpart" of the random field $\theta$. In the zeroth
approximation the functional (\ref{Gamma1}) coincides with $F[n,p]$
and in higher orders the kernel $\Gamma_{n,p}(x;x_1,\ldots,x_n)$ is
given by the sum of diagrams shown in Fig.\,\ref{fig12} (up to two
loops). The analysis of the diagrams in Fig.\,\ref{fig12} shows that
for each diagram and for any argument $x_i$, the corresponding
spatial derivative can be isolated as an external factor. Therefore,
using integration by parts, it is appropriate to move them onto the
corresponding fields $\theta(x_i)$ in Eq.\,(\ref{Gamma1}). As a
result the functional (\ref{Gamma1}) takes the form
\begin{eqnarray}
\Gamma_{n,p}(x;\theta)&=& \frac{1}{n!}\int dx_{1}\cdots\int
dx_{n}\,a_{i_1}(x_{1})\cdots a_{i_n}(x_{n}) \nonumber \\
&& \times \Gamma^{\prime i_1\ldots
i_n}_{n,p}(x;x_1,\ldots,x_n),\label{Gamma2}
\end{eqnarray}
where we define new vector fields $a_{i}(x)=\partial_i \theta(x)$.

\input epsf

The black circles on the top of the diagrams in Fig.\,\ref{fig12}
represents the vertex of the composite operator $F[n,p]$.
Analytically it corresponds to the following expression:
\begin{equation}
V_k(x;x_1,\ldots,x_k)=\frac{\delta^k F[n,p]}{\delta
\theta(x_1)\ldots\delta\theta(x_k)},\label{vertexx}
\end{equation}
where $k$ denotes the number of attached lines. It is possible to
represent it in the following convenient form (see
Ref.\,\cite{AdAnBaKaVa01}):
\begin{equation}
V(x;\,x_{1},\ldots,x_{k})=\prod_{j=1}^k\partial_{i_j}\delta(x-x_{j})\frac{\partial^{k}
F[n,p]}{\partial a_{i_1}\ldots\partial a_{i_k}},\label{Vertex}
\end{equation}
where $a_{i}$ is replaced by
$\partial_{i}\theta(x)$
after the differentiation.

For example, the one-loop diagram shown in Fig.\,\ref{fig12} has
then the following analytical form:
\begin{eqnarray}
K_1 &=& \int dx_{1}\cdots\int dx_{4}V(x;\, x_{1},x_{2})
\langle\theta(x_{1})\theta'(x_{3})\rangle_{0} \nonumber
\\ \times &&  \hspace{-0.7cm} \langle\theta(x_{2})\theta'(x_{4})\rangle_{0}\langle
v_{k}(x_{3})v_{l}(x_{4})\rangle_{0}\partial_{k}\theta(x_{3})\partial_{l}\theta(x_{4}),\label{Coor}
\end{eqnarray}
and analogously one can write down all two-loop diagrams in
Fig.\,\ref{fig12}.

To determine the renormalization constants $Z^{-1}_{[n,p]}$ it is
enough to calculate the function $\Gamma^{\prime i_1\ldots
i_n}_{n,p}(x;x_1,\ldots,x_n)$ with appropriate choice of its
arguments $a_i$ since the function $\Gamma^{\prime i_1\ldots
i_n}_{n,p}(x;x_1,\ldots,x_n)$ contains only logarithmic divergences.
A sufficient choice is to replace them by the fixed point $x$, the
argument of the operator $F[n,p]$. Thus the expression
$a_{i_1}(x)\cdots a_{i_n}(x)$ in (\ref{Gamma2}) can be taken outside
the integration. As a result we come to the local composite operator
$\Gamma_{n,p}(x;\theta)$:
\begin{eqnarray}
\Gamma_{n,p}(x;\theta)&=& \frac{1}{n!}a_{i_1}(x)\cdots a_{i_n}(x) \label{Gamma3} \\
& \times & \int dx_{1}\cdots\int dx_{n}\, \Gamma^{\prime i_1\ldots
i_n}_{n,p}(x;x_1,\ldots,x_n).\nonumber
\end{eqnarray}
After integration one obtains an expression independent of
coordinates. The vector indices of $\Gamma^{\prime i_1\ldots
i_n}_{n,p}$ are transformed into combinations of Kronecker $\delta$
symbols and their contractions with vector symbols of the product
$a_{i_1}(x)\cdots a_{i_n}(x)$ give the original composite operator
$F[n,p]$. Expression (\ref{Coor}) can then be written, up to an
uv-finite part, in the form
\begin{equation} K_1= a_{k}a_{l}\,\frac{\partial^{2}}{\partial
a_{i}\partial a_{j}} F[n,p] X^{K_1}_{ij,\, kl},\label{Coor2}
\end{equation}
with
\begin{eqnarray} X^{K_1}_{ij,\, kl} &=&\int dx_{3}\,\int
dx_{4}\,\,\partial_{i}\langle\theta(x)\theta'(x_{3})\rangle_{0}
\,\partial_{j}\langle\theta(x)\theta'(x_{4})\rangle_{0} \nonumber
\\ && \times \langle v_{k}(x_{3})v_{l}(x_{4})\rangle_{0}.\label{X}
\end{eqnarray}

Two-loop diagrams (Fig.\,\ref{fig12}) can be given in the same form,
namely,
\begin{eqnarray}
K_{2z} &=& a_{k_1}a_{k_2}\,\frac{\partial^{2}F[n,p]}{\partial
a_{i_1}\partial a_{i_2}}  X^{K_{2z}}_{i_1 i_2,\, k_1 k_2}, \nonumber \\
K_{2e} &=& a_{k_1}a_{k_2}a_{k_3}\,\frac{\partial^{3}F[n,p]}{\partial
a_{i_1}\partial a_{i_2}\partial a_{i_3}}
X^{K_{2e}}_{i_1i_2i_3,\,k_1k_2k_3},   \label{KKK}\\
K_{2f} &=&
a_{k_1}a_{k_2}a_{k_3}a_{k_4}\,\frac{\partial^{4}F[n,p]}{\partial
a_{i_1}\partial a_{i_2}\partial a_{i_3}\partial a_{i_4}}
X^{K_{2f}}_{i_1i_2i_3i_4,\,k_1k_2k_3k_4}, \nonumber
\end{eqnarray}
where $z=a,b,c,d$. The analytical expressions for $X^{K_{2z}}$
($z=a,b,c,d,e,f$) can be easily written in analogy with the one-loop
diagram $K_1$; therefore we shall not give their explicit form here.
The tensors $X^{K_{2z}}$ in Eq.\,(\ref{KKK}) can be decomposed into
basic structures made of Kronecker $\delta$ symbols.
\begin{eqnarray}
X^{K_{1}}_{i_1i_2, k_1k_2}&=& \sum_{j=1}^2 A^{K_1}_j
T^{(j)}_{i_1i_2,k_1k_2},\label{X1a}\\
X^{K_{2z}}_{i_1i_2, k_1k_2}&=& \sum_{j=1}^2 A^{K_{2z}}_j
T^{(j)}_{i_1i_2,k_1k_2},\label{X1b} \\
X^{K_{2e}}_{i_1i_2i_3, k_1k_2k_3}&=& \sum_{j=1}^2 A^{K_{2e}}_j
T^{(j)}_{i_1i_2i_3,k_1k_2k_3},\label{X1c} \\
X^{K_{2f}}_{i_1i_2i_3i_4, k_1k_2k_3k_4}&=& \sum_{j=1}^3 A^{K_{2f}}_j
T^{(j)}_{i_1i_2i_3i_4,k_1k_2k_3k_4},\label{X1d}
\end{eqnarray}
where $z=a,b,c,d$, and the tensor structures are defined as
\begin{equation}
T^{(1)}_{ij,kl}= \delta_{ij}\delta_{kl},\,\,\,
T^{(2)}_{ij,kl}=\frac{
\delta_{ik}\delta_{jl}+\delta_{il}\delta_{jk}}{2},
\end{equation}
\begin{eqnarray}
T^{(1)}_{ijk,lmn}&=&
\frac{1}{9}\Big(\delta_{ij}(\delta_{kl}\delta_{mn}+
\delta_{km}\delta_{ln}+\delta_{kn}\delta_{lm}) \nonumber
\\ && \hspace{0.5cm} + (j \leftrightarrow k)+(i \leftrightarrow
k)\Big),
\end{eqnarray}
\begin{eqnarray}
T^{(2)}_{ijk,lmn}&=&
\frac{1}{6}\Big(\delta_{il}(\delta_{jm}\delta_{kn}+
\delta_{jn}\delta_{km}) \nonumber
\\ && \hspace{0.5cm} + (l \leftrightarrow m)+(l \leftrightarrow
n)\Big),
\end{eqnarray}
\begin{eqnarray}
T^{(1)}_{ijkl,mnop}&=&\frac{1}{9}(\delta_{ij}\delta_{kl}+
\delta_{ik}\delta_{jl}+\delta_{il}\delta_{jk}) \nonumber \\
&& \times (\delta_{mn}\delta_{op}+
\delta_{mo}\delta_{np}+\delta_{mp}\delta_{on}),
\end{eqnarray}
\begin{eqnarray}
T^{(2)}_{ijkl,mnop}&=&\frac{1}{72} \nonumber
\\ \times \,\,\,\,\,\,\,\,\,\,\,\,\,\,\,\,\,\,\,\, && \hspace{-1.8cm}
 \Big(\delta_{ij}\big(\delta_{mn}(\delta_{ko}\delta_{lp}+\delta_{kp}\delta_{lo})+
 \delta_{mo}(\delta_{kn}\delta_{lp}+\delta_{kp}\delta_{ln})\nonumber
 \\ + \,\,\,\,\,\,\,\,\, && \hspace{-1.0cm}
 \delta_{mp}(\delta_{kn}\delta_{lo}+\delta_{ko}\delta_{ln})+
 \delta_{no}(\delta_{km}\delta_{lp}+\delta_{kp}\delta_{lm})
 \nonumber \\ + \,\,\,\,\,\,\,\,\, && \hspace{-1.0cm}
 \delta_{np}(\delta_{km}\delta_{lo}+\delta_{ko}\delta_{lm})+
 \delta_{op}(\delta_{km}\delta_{ln}+\delta_{kn}\delta_{lm})
\big) \nonumber \\
+ \,\,\,\,\,\,\,\,\,\,\, && \hspace{-1.2cm} (j \leftrightarrow k) +
(j \leftrightarrow l) + (i \leftrightarrow k) + (i \leftrightarrow
l) \nonumber \\ + \,\,\,\,\,\,\,\,\,\,\, && \hspace{-1.2cm} (i
\leftrightarrow k, j\leftrightarrow l) \Big),
\end{eqnarray}
\begin{eqnarray}
T^{(3)}_{ijkl,mnop}&=&\frac{1}{24} \nonumber
\\ \times \,\,\,\,\,\,\,\,\,\,\,\,\,\,\,\,\,\,\,\, && \hspace{-1.8cm}
 \Big(\delta_{im}\big(\delta_{jn}(\delta_{ko}\delta_{lp}+\delta_{kp}\delta_{lo})+
 \delta_{jo}(\delta_{kn}\delta_{lp}+\delta_{kp}\delta_{ln})\nonumber
\\ + \,\,\,\,\,\,\,\,\, && \hspace{-1.0cm}
 \delta_{jp}(\delta_{kn}\delta_{lo}+\delta_{ko}\delta_{ln}) \big)
  \\ + \,\,\,\,\,\,\,\,\,\,\, && \hspace{-1.2cm} (m \leftrightarrow n) +
(m \leftrightarrow o) + (m \leftrightarrow o) + (m \leftrightarrow
p)
 \Big),\nonumber
\end{eqnarray}
and the scalar coefficients $A_j^x$ ($x=K_1,K_{2a},\ldots,K_{2f}$)
in Eqs.\,(\ref{X1a})-(\ref{X1d}) are given as
\begin{eqnarray}
A_1^{x}&=&\frac{(d+1)X_1^{x}-2 X_2^{x}}{d(d+2)(d-1)},\label{A1x}\\
A_2^{x}&=&\frac{2(-X_1^{x}+d X_2^{x})}{d(d+2)(d-1)},
\end{eqnarray}
for $x=K_1,K_{2a},K_{2b},K_{2c},K_{2d}$. Further, for operator
$K_{2e}$ we have
\begin{eqnarray}
A_1^{K_{2e}}&=&\frac{9((d+1)X_1^{K_{2e}}-2 X_2^{K_{2e}})}{d(d+2)(d+4)(d-1)},\\
A_2^{K_{2e}}&=&\frac{6(-3 X_1^{K_{2e}}+(d+2)
X_2^{K_{2e}})}{d(d+2)(d+4)(d-1)}, \label{A2K2e}
\end{eqnarray}
and for operator $K_{2f}$
\begin{widetext}
\begin{eqnarray}
A_1^{K_{2f}}&=&\frac{9(d+3)\left((d+5)X_1^{K_{2f}}-8
X_2^{K_{2f}}\right)+72 X_3^{K_2f}}{(d-1)d(d+1)(d+2)(d+4)(d+6)},\\
A_2^{K_{2f}}&=&-\frac{72\left((d+3)X_1^{K_{2f}}-(d^2+3d+6)X_2^{K_{2f}}
+2(d+2)X_3^{K_{2f}} \right)}{(d-1)d(d+1)(d+2)(d+4)(d+6)}, \\
A_3^{K_{2f}}&=&\frac{24\left(3 X_1^{K_{2f}}+(d+2)\left(-6
X_2^{K_{2f}} +(d+4)X_3^{K_{2f}}\right)
\right)}{(d-1)d(d+1)(d+2)(d+4)(d+6)},
\end{eqnarray}
\end{widetext}
where $X_j^x$, $j=1,2,3, x=K_1,K_{2a},\ldots,K_{2f}$ are given in
Appendix \ref{ApD} (calculations are performed in the MS scheme).
Looking at expressions (\ref{D2}) and (\ref{D16}) in Appendix
\ref{ApD} one can see the explicit dependence of the diagrams
$K_{2a}$ and $K_{2d}$ on the helicity parameter $\rho$. These
diagrams are not present in the case of the rapid-change model at
all ($u\rightarrow\infty$; see Ref.\,\cite{AdAnBaKaVa01}) because
they contain closed circuits of retarded propagators $\langle\theta
\theta^{\prime}\rangle_0$ and therefore automatically vanish (for
the same reason the self-energy operator
$\Sigma_{\theta\theta^{\prime}}$ of the rapid-change model has only
one-loop corrections \cite{AdAnBaKaVa01}). This is one of the
important reasons to study higher-loop corrections of models with
finite correlation time, namely, some considerable properties cannot
be studied within simple rapid-change models.

Let us briefly concentrate our attention on the comparison of the
rapid-change limit of our two-loop results for composite operators
with those obtained in Ref.\,\cite{AdAnBaKaVa01}. This comparison
leads to some nontrivial results for the corresponding integrals.
They are presented in Appendix \ref{ApC} [expressions
(\ref{ApC4}),(\ref{ApC5}), and (\ref{ApC6})]. We found one misprint
in Eq.\,(5.42) of Ref.\,\cite{AdAnBaKaVa01}, namely, there must be
an overall factor $d^2-1$ in the expression for $A_1$.

The critical dimensions of our operators are defined by th egeneral
formula given in Eq.\,(\ref{deltaa}). When we rewrite it in the
concrete form of the operator $F_{np}\equiv F[n,p]=
\partial_{i_1}\theta \cdots \partial_{i_p}\theta (\partial_{i}\theta
\partial_{i} \theta)^l$ then we have
\begin{equation}
\Delta_{F_{np}}=d_{F_{np}}^k+\Delta_{\omega}d_{F_{np}}^{\omega}+\gamma^*_{F_{np}}\,.\label{deltaaa}
\end{equation}
Now using the canonical dimensions shown in Table \ref{table1},
namely, $d^k_{\theta}=-1$ and $d^{\omega}_{\theta}=0$ one
immediately comes to the result
\begin{equation}
\Delta_{F_{np}}=\gamma^*_{F_{np}}\,.\label{deltaaaa}
\end{equation}
This means that the critical dimensions of our operators are equal
to the corresponding anomalous dimensions at a corresponding fixed
point.

The first step to determine the anomalous dimensions is to calculate
the constants $Z_{np}\equiv Z_{[n,p][n,p]}$ [see
Eq.\,(\ref{anomdimF1})] in the two-loop approximation. In our case
it is given as
\begin{equation}
Z_{np}=1+\frac{g}{\varepsilon} A_{np} + \frac{g^2}{\varepsilon}
B_{np} + \frac{g^2}{\varepsilon^2} C_{np}\,. \label{zall}
\end{equation}
The coefficient $C_{np}$ will not contribute into the corresponding
anomalous dimension (this can be verified by direct calculation);
hence we do not present its explicit form in what follows. The
coefficients $A_{np}$ and $B_{np}$ are defined as
\begin{equation}
A_{np}=\frac12\left(k_1^{(2)} A_1^{K_1}+ k_2^{(2)}
A_2^{K_1}\right)\, \label{Anp}
\end{equation}
and
\begin{eqnarray}
B_{np}&=&k_1^{(2)}\left(\frac12 A_1^{K_{2a}} +\frac12 A_1^{K_{2a}} +
A_1^{K_{2c}} + A_1^{K_{2d}} \right) \nonumber \\
&+&  k_2^{(2)}\left(\frac12 A_2^{K_{2a}} +\frac12 A_2^{K_{2a}} +
A_2^{K_{2c}} + A_2^{K_{2d}} \right) \nonumber \\ &+&  k_1^{(3)}
A_1^{K_{2e}}+ k_2^{(3)} A_2^{K_{2e}}\,, \label{Bnp}
\end{eqnarray}
where $A_x^y, x=\{1,2\}, y=\{K_1, K_{2z}\}, z=\{a,b,c,d,e\}$ are
defined in Eqs.\,(\ref{A1x})-(\ref{A2K2e}) and
\begin{eqnarray}
k_1^{(2)}&=&(n-p)(d+n+p-2)\,, \\
k_2^{(2)}&=&n(n-1)\,, \\
k_1^{(3)}&=&(n-2)(n-p)(d+n+p-2)\,, \\
k_2^{(3)}&=&n(n-1)(n-2).
\end{eqnarray}

Then the anomalous dimensions have the form
\begin{equation}
\gamma_{np}\equiv \gamma_{F_{np}}=-2 A_{np}\,\, g -4 B_{np}\,\,
g^2\,. \label{gammakon}
\end{equation}
Thus, the coefficient $A_{np}$ represents the one-loop contribution
to the anomalous dimension, and the coefficient $B_{np}$ the
two-loop one. The critical dimension $\Delta_{F_{np}}$ [see
Eq.\,(\ref{deltaaaa})] of the operator $F_{np}$ is obtained from
(\ref{gammakon}) when it is taken at the corresponding fixed point.

\subsection{ Anomalous scaling: Two-loop
approximation}

Our aim is the investigation of the influence of the helicity on the
anomalous scaling in the most interesting situation of the
degenerate fixed point, namely, the fixed point denoted as FPV in
Sec.\ref{sec:ScalReg}. In this case, the dimensions
$\Delta_{F_{np}}$ are represented in the following series in the
only independent exponent $\varepsilon=\eta$ [it is obtained from
(\ref{gammakon}) by the substitution of the corresponding fixed
point for $g_*$]
\begin{equation}
\Delta_{F_{np}}=\varepsilon \Delta^{(1)}_{F_{np}} + \varepsilon^2
\Delta^{(2)}_{F_{np}}\,.
\end{equation}
The one-loop contribution has the form
\begin{equation}
\Delta^{(1)}_{F_{np}}=\frac{2n(n-1)-(n-p)(d+n+p-2)(d+1)}{2(d+2)(d-1)}\,,
\end{equation}
which is independent of the parameter u (the ratio of the velocity
correlation time and the scalar turnover time). Although the fixed
point value $g_*$ given by Eq.\,(\ref{fixh}) and the coefficient
$B_{np}$ in Eq.\,(\ref{gammakon}) explicitly depend on the helicity
parameter $\rho$, the two-loop contribution to the critical
dimension $\Delta^{(2)}_{F_{np}}$ is independent of $\rho$. Thus,
the result is the same as that obtained in Ref.\,\cite{AdAnHo02}
(there is a misprint in the final explicit result but the correct
formula was republished in Ref.\,\cite{AdAnHoKi05}). Its explicit
expression is rather large, and as it can be found elsewhere we
shall not repeat it here. At first sight this result is a surprise
but it can be understood in the following, rather simple, way. As we
know the structure functions $S_{n}(r)$ (which are studied here) are
functions of the value of the distance $r=|{\bf x}-{\bf
x^{\prime}}|$. Therefore, only those phenomena will have impact on
the critical dimensions that can "change" the spatial distances.
Among such phenomena belong the compressibility and anisotropy. As
for helicity, it breaks the mirror symmetry but it does not disturb
spatial distances. Therefore, it cannot influence the critical
dimensions, i.e., it cannot change the corresponding asymptotic
behavior. Thus, if our statement is right then we expect that the
situation will be the same in all orders of perturbation expansion,
namely, the quantities such as effective diffusivity will depend on
helicity, but critical dimensions of the structure functions will
not. But, of course, for now it is only a speculation and the
independence of the critical dimensions of helicity is maybe only
the effect of two-loop approximation. To solve this problem at least
three-loop calculations are needed.

On the other hand, to study the helicity effect on the two-loop
level it is enough to avoid the conditions of isotropy or
incompressibility of the system. Thus, the next step is, e.g., to
include the assumption of compressibility of the system and
investigate the combined effects of the helicity and compressibility
on the scaling properties of the model under consideration. We
assume that a nontrivial result can be obtained.

\begin{figure}[t]
     \vspace{-1.65cm}
       \begin{center}
       \leavevmode
       \epsfxsize=8cm
       \epsffile{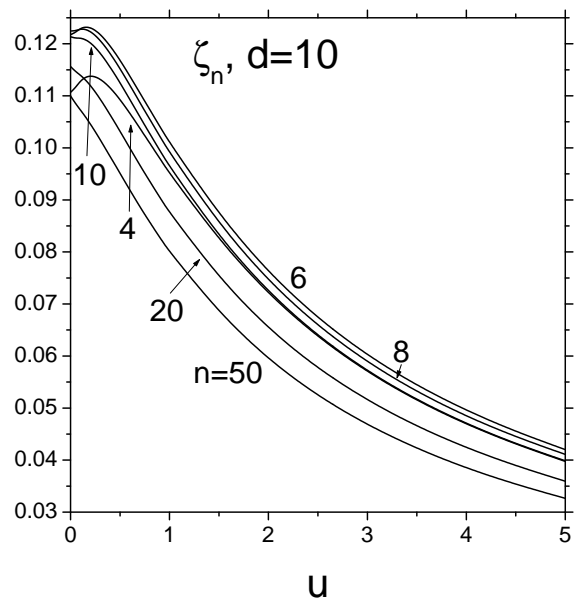}
   \end{center}
\vspace{-1cm} \caption{Behavior of the quantity $\zeta_n $ (see
text) for $n=4,6,8,10,20,$ and $50$ as a function of $u$ for $d=10$
in units of $10^{-3}$. In Ref.\,\cite{AdAnHo02} the curves are
interpreted as $n=4,6,8,20$ (from bottom to top). But in fact they
correspond to $n=20,4,8,6$ (from bottom to top). The curves for
$n=10, 50$ are added to demonstrate the situation more completely.
\label{fig13}}
\end{figure}

As was already mentioned detailed analysis of the two-loop
contribution to the critical dimensions of the structure functions
within the model under our consideration (without helicity) was done
in Ref.\,\cite{AdAnHo02}. We have recalculated their results and
found some discrepancy in interpretations of our and their numerical
results. That is, as our calculations show, a hierarchical behavior
of the quantity $\zeta_n \equiv
\left[\Delta_{n0}^{(2)}-\Delta_{n0}^{(2)}|_{u=\infty}\right]/n^3$ as
a function of $n$ for a concrete value of $d$ (dimension of space)
is destroyed in Fig.\,1(d) in Ref.\,\cite{AdAnHo02}.  This figure
corresponds to a large enough value of $d$ (namely, $d=10$). Our
calculations lead to the same curves as theirs but they correspond
to different values of $n$ which can be seen by direct comparison of
Fig.\,1(d) in Ref.\,\cite{AdAnHo02} and Fig.\,\ref{fig13} in the
present paper.

\input epsf

The conclusion is as follows: The hierarchical behavior of the
quantities $\zeta_n $ is not present for large enough space
dimensions $d$ (the same situation also occurs for other large
values of $d$ as can be shown by direct calculations).

All the other results are the same as in Ref.\,\cite{AdAnHo02}; thus
we shall not repeat them here.

\section{Conclusion}\label{sec:Conc}

In this paper, the influence of helicity on the stability of
asymptotic regimes, on the anomalous scaling, and on the effective
diffusivity was studied in the framework of the passive scalar
advected by the turbulent flow with finite correlations in time of
the velocity field. Such investigation is important and useful for
understanding of efficiency of toy models (like the Kraichnan and
related model) to study the real turbulent motions by means of
modern theoretical methods including renormalization group approach.
Thus, it can be considered as the first step in investigation of the
influence of helicity in a real turbulent environment.

In the present paper, the RG calculations are performed in the
two-loop approximation, which is necessary to include the effects of
helicity. It was shown that the anomalous scaling of the structure
functions, which is typical for the Kraichnan model and its numerous
extensions (see the Introduction), is not changed by the inclusion
of helicity in the incompressible fluid. This is given
mathematically by the very interesting fact that although separated
two-loop Feynman diagrams of the corresponding composite operators
strongly depend on the helicity parameter $\rho$, their sum - the
critical dimension $\Delta_{n}$ - is independent of $\rho$ in the
asymptotic regime defined by an ir stable fixed point. This very
interesting fact can be explained physically by rather simple
arguments in the following way (as was discussed in the previous
section maybe it is only the effect of the two-loop approximation,
therefore, to confirm what follows higher-loop calculations are
needed). The single-time structure functions $S_{n}(r)$ of the
scalar field depend only on the spatial distance $r=|{\bf x}-{\bf
x^{\prime}}|$ but not on the direction. Thus, we suppose that only
this phenomenon will change the critical dimensions of structure
functions which modify spatial relations. It can be, e.g., the
inclusion of compressibility or spatial anisotropy. On the other
hand, helicity breaks the mirror symmetry, which is not related to
distance. As a result, the critical dimensions of the structure
functions are not affected by the helicity of the system. We suppose
that an analogous situation will hold for all quantities that depend
only on the spatial distance (more precisely, that are constructed
from quantities taken at different spatial points), and in all
orders of perturbation theory. The situation can be different when
one includes in the investigation the effects of helicity together
with another assumption about the turbulent flow, e.g., its
compressibility. We suppose that nontrivial results can be obtained
in this more general case.

On the other hand, the stability of possible asymptotic regimes, the
values of the fixed RG points, and the turbulent diffusivity
strongly depend on the amount of helicity. It is shown that the
presence of helicity in the system leads to restrictions on the
possible values of the parameters of the model. The most interesting
fact is the existence of a critical value $\rho_{c}$ of the helicity
parameter $\rho$ which divides the interval of possible absolute
values of $\rho$ into two parts with completely different behavior.
It is related to the existence of a competition between nonhelical
and helical contributions within the two-loop approximation. As a
result of this competition, within the so-called frozen limit, the
presence of helicity enlarges the region of parameter space with a
stable scaling regime, and if $|\rho|=\rho_{c}$ the corresponding
two-loop restriction  vanishes completely and one comes to the
one-loop results \cite{Antonov99}. Similar splitting, although more
complicated, into two nontrivial behaviors of the fixed point was
also obtained in the general case with finite correlations in time
of the velocity field.

Another quantity which rather strongly depends on the helicity
parameter $\rho$ is the effective diffusivity. It is shown that the
value of effective diffusivity can be 50\% larger in the helical
case in comparison with the nonhelical case.

\begin{acknowledgments}
M.H. is thankful to N.V.Antonov and L.Ts. Adzhemyan for discussion.
The work was supported in part by VEGA Grant No. 6193 of the Slovak
Academy of Sciences, by the Science and Technology Assistance Agency
under Contract No. APVT-51-027904, by  RFFI - RFBR Grants No.
05-05-64735 and 05-02-17603, and by COFIN "Sistemi Complessi e
Problemi a Molti Corpi".
\end{acknowledgments}

\appendix

\section{}\label{ApA}

In principle a few ways exist to evaluate the two-loop diagrams
$B_1$ and $B_2$ which are explicitly shown in Fig.\,\ref{fig3}. We
compare two of them on the example of the Feynman diagram $B_1$. The
explicit analytical expression for $B_1$ in the
wave-number-frequency representation is
\begin{eqnarray}
B_1&=&\frac{g^2\nu^6\mu^{4\varepsilon}}{(2\pi)^{2d+2}} \int
\frac{d^d {\bf
k} d^d {\bf q} d\omega_k d\omega_q k^{4-d-2\varepsilon} q^{4-d-2\varepsilon}}
{(\omega^2_k+\nu^2 u^2 k^4)(\omega^2_q+\nu^2 u^2 q^4)} \nonumber\\
& & \hspace{-1.5cm}\times \frac{
p_{i_1}(p-k)_{i_3}(p-k-q)_{i_4}(p-k)_{i_2} P^{\rho}_{i_1
i_2}(k)P^{\rho}_{i_3 i_4}(q)}{(i \omega_k+\nu({\bf p}-{\bf
k})^2)^2(i(\omega_k+\omega_q)+\nu({\bf p}-{\bf k}-{\bf q})^2)},
\end{eqnarray}
where ${\bf p}$ denotes the external wave number (momentum), ${\bf
k}$ and ${\bf q}$ are two independent internal wave numbers, the
external frequency $\omega_p$ is taken to be zero (we are only
interested in the divergent part of the diagram and it is
independent of $\omega_p$), $\omega_k$ and $\omega_q$ are the
corresponding internal frequencies, $\eta$ is taken to be zero (see
discussion in Sec.\,\ref{sec:RG}), $P^{\rho}_{ij}$ is the helical
transverse projector defined in Sec.\,\ref{sec:Model}, and over the
internal vector indices $i_j, j=1,2,3,4$, corresponding summations
are assumed.

After integration over the internal frequencies, which is rather
simple, and then making an expansion in respect of the external
momentum ${\bf p}$ and leaving only terms of order two in respect of
${\bf p}$ (because uv divergences can have only this structure; see
Sec.\,\ref{sec:RG}), and, in the end, after summation over vector
indices one has
\begin{eqnarray}
B_1&=&\frac{g^2\nu \mu^{4\varepsilon}}{(2\pi)^{2d} 4 u^2(1+u)^2}
\int \frac{d^d {\bf k} d^d {\bf q} k^{-d-2\varepsilon}
q^{-d-2\varepsilon}} {((1+u)(k^2+q^2)+2{\bf k} \cdot {\bf q})} \nonumber\\
& & \hspace{-1.25cm}\times \Bigg[p^2q^2-\frac{({\bf p}\cdot{\bf
k})^2q^2}{k^2}-\frac{p^2({\bf k}\cdot{\bf q})^2}{k^2}+\frac{({\bf
p}\cdot{\bf k})^2({\bf k}\cdot{\bf
q})^2}{k^4}\Bigg]\hspace{-0.08cm}. \label{B1aaa}
\end{eqnarray}
Now we shall calculate $B_1$ in two different ways.

The first approach to the integral $B_1$ is based on division of
integrations into radial and  angle parts which after standard
procedures leads to (for details see, e.g.,
Ref.\,\cite{AdAnBaKaVa01})
\begin{eqnarray}
B_1&=&\frac{g^2\nu \mu^{4\varepsilon} p^2 S_d S_{d-1}}{(2\pi)^{2d} 4
u^2(1+u)^2} \frac{d-1}{d} \label{B1aa} \\
& & \hspace{-1.2cm} \times \int_m^\infty \frac{d
k}{k^{1+2\varepsilon}} \int_m^\infty \frac{d q}{q^{1+2\varepsilon}}
\int_{-1}^1 d x \frac{(1-x^2)^{(d-1)/2} q^2}{(1+u)(k^2+q^2)+2 k q
x},\nonumber
\end{eqnarray}
where $x=\cos\psi$, and $\psi$ denotes the angle between vectors
${\bf k}$ and ${\bf q}$. $S_d$ and $S_{d-1}$ are $d$-dimensional and
$(d-1)$-dimensional sphere, respectively, which are defined as
$S_d=2\pi^{d/2}/\Gamma(d/2)$. The needed ir cutoff of the
integrations is represented by $m \sim L^{-1}$. It is useful to
rewrite the denominator of the expression under the integration in
the form of the infinite series
\begin{equation}
\frac{1}{(1+u)(k^2+q^2)+2 k q x}=\sum_{j=1}^\infty \frac{(-2 k q
x)^{j-1}}{[(k^2+q^2)(1+u)]^{j}}
\end{equation}
Now we are able to integrate over the angle variable $x=\cos\psi$
term by term in the sum, which yields
\begin{eqnarray}
B_1&=&\frac{g^2\nu \mu^{4\varepsilon} p^2 S_d S_{d-1}}{(2\pi)^{2d} 8
u^2(1+u)^2} \frac{d-1}{d} \Gamma\left(\frac{d+1}{2}\right) \nonumber
\\ & & \sum_{j=1}^\infty ((-1)^{j-1}+1)
\frac{\Gamma\left(\frac{j}{2}\right)}{\Gamma\left(\frac{d+j+1}{2}\right)}
\nonumber \\
& & \hspace{-0.3cm} \int_m^\infty \frac{d k}{k^{1+2\varepsilon}}
\int_m^\infty \frac{d q}{q^{1+2\varepsilon}}  \frac{q^2 (2 k
q)^{j-1}}{[(k^2+q^2)(1+u)]^{j}}.\label{b1bbb}
\end{eqnarray}
The factor $[(-1)^{j-1}+1]$ keeps only odd terms of the series.
Therefore, we can redefine the summation in Eq.\,(\ref{b1bbb}) in
the following form:
\begin{eqnarray}
B_1&=&\frac{g^2\nu \mu^{4\varepsilon} p^2 S_d
S_{d-1}(d-1)\Gamma\left(\frac{d+1}{2}\right)}{(2\pi)^{2d} 4
u^2(1+u)^2 d} \sum_{j=0}^\infty
\frac{\Gamma\left(\frac{2j+1}{2}\right)}{\Gamma\left(\frac{d+2j+2}{2}\right)}
\nonumber \\
& & \hspace{-0.3cm} \int_m^\infty \frac{d k}{k^{1+2\varepsilon}}
\int_m^\infty \frac{d q}{q^{1+2\varepsilon}}  \frac{q^2 (2 k
q)^{2j}}{[(k^2+q^2)(1+u)]^{2j+1}}.\label{b1bbc}
\end{eqnarray}
To proceed it is appropriate to make transformation of  the
variables $k,q$ to the new polar coordinates, namely: $k=s\cos\phi,
q=s\sin\phi$. This gives
\begin{eqnarray}
B_1&=&\frac{g^2\nu \mu^{4\varepsilon} p^2 S_d
S_{d-1}(d-1)\Gamma\left(\frac{d+1}{2}\right)}{(2\pi)^{2d} 4
u^2(1+u)^2 d} \nonumber \\
& &\sum_{j=0}^\infty
\frac{\Gamma\left(\frac{2j+1}{2}\right)}{\Gamma\left(\frac{d+2j+2}{2}\right)}
  \frac{2^{2j}}{(1+u)^{2j+1}}\int_m^\infty
 \frac{d s}{s^{1+4\varepsilon}} \nonumber \\ & & \int_0^{\pi/2} d \phi
\cos\phi^{2j-1-2\varepsilon}
\sin\phi^{2j+1-2\varepsilon}.\label{b1bbd}
\end{eqnarray}
The integral over the radial variable $s$ is trivial, and the
integral over angle $\phi$ can be found, e.g., in
Ref.\,\cite{GradRizi}. Then one has
\begin{eqnarray}
B_1&=&\frac{g^2\nu p^2 \pi^ {1/2} S_d
S_{d-1}(d-1)\Gamma\left(\frac{d+1}{2}\right)}{(2\pi)^{2d} 32
u^2(1+u)^2 d} \left(\frac{\mu}{m}\right)^{4\varepsilon} \frac{1}{\varepsilon} \nonumber \\
& &\hspace{-0.5cm}\sum_{j=0}^\infty
\frac{\Gamma\left(\frac{2j+1}{2}\right)}{\Gamma\left(\frac{d+2j+2}{2}\right)}
\frac{2^{2\varepsilon}}{(1+u)^{2j+1}}
\frac{\Gamma\left(j-\varepsilon\right)}{\Gamma\left(\frac12+j-\varepsilon\right)}.\label{b1bbe}
\end{eqnarray}
In the end, the summation over $j$ leads to the final result for
$B_1$ which is given in Eq.\,(\ref{b2}), where only the divergent
part is shown. $\Gamma\left(j-\varepsilon\right)$ in
Eq.\,(\ref{b1bbe}) for $j=0$ has a pole with respect to
$\varepsilon$, which gives the pole of the second order in $B_1$ in
Eq.\,(\ref{b2}).

The second approach to the calculation of our two-loop diagrams is
as follows. We start with the expression for $B_1$ as represented in
Eq.\,(\ref{B1aaa}). Now using the well-known formula of Feynman
parametrization
\begin{eqnarray}
\frac{1}{A_1^{\alpha_1} A_2^{\alpha_2}\cdots A_n^{\alpha_n}} &=&
\frac{\Gamma\Big(\sum\limits_{i=1}^n \alpha_i
\Big)}{\prod\limits_{i=1}^n \Gamma\left(\alpha_i\right)}  \nonumber
\\ & &\hspace{-3.5cm}\times \int_0^1 \cdots \int_0^1 d u_1\cdots d
u_n\frac{\delta\Big(\sum\limits_{i=1}^n \alpha_i
-1\Big)\prod\limits_{i=1}^n
u_i^{\alpha_i-1}}{\Big(\sum\limits_{i=1}^n A_i
u_i\Big)^{\sum\limits_{i=1}^n \alpha_i}}
\end{eqnarray}
$B_1$ obtains the form
\begin{eqnarray}
B_1&=&\frac{g^2\nu \mu^{4\varepsilon}
\Big(\frac{d}{2}+\varepsilon\Big)}{(2\pi)^{2d} 4 u^2(1+u)^2}
\int_0^1 d u_1 u_1^{\frac{d}{2}+\varepsilon-1}\nonumber \\
& & \hspace{-0.5cm}\int \frac{d {\bf k} d {\bf q}
\left(p^2-\frac{({\bf p}\cdot{\bf
k})^2}{k^2}\right)\left(q^2-\frac{({\bf k}\cdot{\bf
q})^2}{k^2}\right)}{k^{d+2 \varepsilon}[X q^2+2Y {\bf k}\cdot{\bf
q}+Z k^2]^{\frac{d}{2}+\varepsilon+1}},
\end{eqnarray}
where
\begin{eqnarray}
X&=& u_1+(1-u_1)(1+u),\nonumber \\
Y&=& 1-u_1, \\Z&=& (1+u)(1-u_1). \nonumber
\end{eqnarray}
The integration over ${\bf q}$ is now done using the general formula
(\ref{formula}) given in Appendix\,\ref{ApB}, namely,
\begin{eqnarray}
\int d {\bf q}\frac{q_i q_j}{[X q^2+2Y {\bf k}\cdot{\bf q}+Z
k^2]^{\frac{d}{2}+\varepsilon+1}} &=&
\frac{\pi^{\frac{d}{2}}X^{-1-\frac{d}{2}}}{\Gamma\left(\frac{d}{2}+\varepsilon+1\right)}\nonumber
\\
& &
\hspace{-6cm}\frac{1}{k^{2\varepsilon}\left(Z-\frac{Y^2}{X}\right)^{\varepsilon}}
\left[\Gamma(\varepsilon)\frac{\delta_{ij}}{2}+\frac{\Gamma(\varepsilon+1)
Y^2}{\left(X Z-Y^2\right)}\frac{k_i k_j}{k^2}\right],
\end{eqnarray}
which yields
\begin{eqnarray}
B_1 &=& \frac{g^2 \nu  \mu^{4\varepsilon} \pi^{\frac{d}{2}}
(d-1)}{(2\pi)^{2d} 8 u^2 (1+u)^2 }
\frac{\Gamma\left(\varepsilon\right)}{\Gamma\left(\frac{d}{2}+\varepsilon\right)}
\nonumber \\
& & \hspace{-1cm} \int_0^1 d u_1
\frac{u_1^{\frac{d}{2}+\varepsilon-1}
X^{-\frac{d}{2}-1}}{\left(Z-\frac{Y^2}{X}\right)^{\varepsilon}} \int
d {\bf k} \frac{\left(p^2-\frac{{\bf p}\cdot{\bf
k}}{k^2}\right)}{k^{d+4\varepsilon}}.
\end{eqnarray}
The integration over vector ${\bf k}$ is now straightforward, after
which we have
\begin{eqnarray}
B_1 &=& \frac{g^2 \nu S_d \pi^{\frac{d}{2}} p^2 (d-1)^2}{(2\pi)^{2d}
32 u^2 (1+u)^2 d \Gamma\left(\frac{d}{2}+\varepsilon\right)}
\left(\frac{\mu}{m}\right)^{4\varepsilon}
\frac{\Gamma(\varepsilon)}{\varepsilon}\nonumber \\
& & \int_0^1 d u_1 \frac{u_1^{\frac{d}{2}+\varepsilon-1}
X^{-\frac{d}{2}-1}}{\left(Z-\frac{Y^2}{X}\right)^{\varepsilon}}\,,
\label{b1ccc}
\end{eqnarray}
where $m \sim L^{-1}$ represents the needed ir regularization. We
are interested only in the divergent (poles in $\varepsilon$) part
of the expression (\ref{b1ccc}). After doing all needed expansions
with respect to $\varepsilon$ one has the final result for the
Feynman diagram $B_1$:
\begin{eqnarray}
B_1&=&\frac{S_d^2 p^2 g^2 \nu (d-1)^2}{(2\pi)^{2d} 32 u^2 (1+u)^3
d^2}\left(\frac{\mu}{m}\right)^{4\varepsilon} \bigg\{
\frac{1}{\varepsilon^2} \nonumber \\
& & \hspace{-1.3cm} + \frac{1}{\varepsilon}\bigg[
\psi(1)-\psi\left(d/2\right) -\frac{2}{d (1+u)^{\frac{d}{2}}}
{_2F_1}\Big(\frac{d}{2},\frac{d}{2};1+\frac{d}{2};\frac{u}{1+u}\Big)
\nonumber \\ & &  - \frac{d(1+u)}{2}\int_0^1 d u_1
\frac{u_1^{\frac{d}{2}-1}
\ln\left(Z-\frac{Y^2}{X}\right)}{X^{\frac{d}{2}+1}}\bigg]
\bigg\},\label{b2a}
\end{eqnarray}
where $\psi(n)$ is the logarithmic derivative of the gamma function,
given by $\psi(n)=\Gamma^{\prime}(n)/\Gamma(n)$.

Thus, we have two different analytical representations of the same
Feynman diagram $B_1$, namely, Eqs.\,(\ref{b2}) and (\ref{b2a}). The
comparison of them leads to the nontrivial result for integral in
Eq.\,(\ref{b2a}), which is present in Appendix \ref{ApC} in
Eq.\,(\ref{int1}). The comparison of the results for diagram $B_2$
(see Fig.\,\ref{fig3}) obtained by these two methods yields other
nontrivial results for some integrals. They are shown in Appendix
\ref{ApC} in Eqs.\,(\ref{int2}) and (\ref{int3}).

\section{}\label{ApB}

In this appendix we introduce the general integral formula which was
used in the previous appendix.

\underline{\bf{Theorem:}} Let $\mathcal{V}$ be a $d$-dimensional
Euclidean vector space over the field of real numbers $\mathbb{R}$.
Let $l,n \in \mathbb{N}$ (natural numbers), and $ {\bf k}^{(i)}$ for
$i = 1,2,\cdots,l$ are vectors in $\mathcal{V}$. Then for an
arbitrary $l\times l$ real matrix $v_{js}$ with $\det v\neq0$,
arbitrary vectors ${\bf a}^{(i)}$ ($i = 1,2,\cdots,l$), and
arbitrary $c,\alpha \in \mathbb{R}$ the following general formula
holds:
\begin{widetext}
\begin{eqnarray}
 \int^{\infty}_{-\infty} \hspace{-10mm} && \ldots
\int^{\infty}_{-\infty} \frac{d {\bf k}^{(1)} \ldots d {\bf k}^{(l)}
k_{i_1}^{(q_{1})}k_{i_{2}}^{(q_{2})}
 \ldots k_{i_{n}}^{(q_{n})}}
{\left[v_{is}  {\bf k}^{(i)} \cdot {\bf k}^{(s)} + 2 {\bf a}^{(i)}
\cdot {\bf k}^{(i)} +c  \right]^\alpha} = \nonumber \\
& & \frac{(-1)^n \pi^{\frac{d
 l}{2}}(\mathrm{det}\;v)^{-\frac{d}{2}}}{\Gamma(\alpha)}\;
\sum_{p=0}^{\lfloor\frac{n}{2}\rfloor} \; \frac{ \Gamma\left(\alpha
-\frac{d l}{2} - \lfloor \frac{n}{2}\rfloor + p \right)\left[ c -
(v^{-1})_{is} {\bf a}^{(i)} \cdot {\bf a}^{(s)} \right]^{\lfloor
\frac{n}{2}\rfloor +
 \frac{d l}{2} - \alpha - p}}{
(\lfloor \frac{n}{2}\rfloor - p)!\,(2 p + n(\mathrm{mod} \;2))!
4^{\lfloor \frac{n}{2}\rfloor -p} } \nonumber \\ & &
 \sum_{ P(q_1 i_1, q_2 i_2,\ldots ,q_n i_n)}
\hspace{-4mm}(v^{-1})_{q_1 s_1} a_{i_{1}}^{(s_{1})}
 (v^{-1})_{q_2 s_2} a_{i_2}^{(s_2)} 
\ldots
 (v^{-1})_{q_{2 p + n(\mathrm{mod}\;2)}s_{2 p + n(\mathrm{mod}\;2)}}
 a_{i_{2p +n(\mathrm{mod}\;2)}}^{(s_{i_{2p +
 n(\mathrm{mod}\;2)}})} \nonumber \\[3mm]
&\times& \delta_{i_{2p + n(\mathrm{mod}\;2)+1} i_{2p +
n(\mathrm{mod}\;2)+2}}(v^{-1})_{q_{2p + n(\mathrm{mod}\;2)+1} q_{2p
+ n(\mathrm{mod}\;2)+2}} \ldots \delta_{i_{n-1}i_{n}}
(v^{-1})_{q_{n-1}q_{n}}, \label{formula}
\end{eqnarray}
\end{widetext}
where the summation is taken over all permutations of the indices
$i_1,i_2,\ldots,i_n$, $k_{j}^{(s)}$ and $a_{j}^{(s)}$ are the $j$th
components of the vectors ${\bf k}^{(s)}$ and ${\bf a}^{(s)}$,
$\delta_{ij}$ denotes the Kronecker delta, and $\lfloor n/2 \rfloor
= n/2$ for even $n$ and $\lfloor n/2 \rfloor = (n-1)/2$ for odd $n$.
Over all dummy indices the corresponding summation is assumed.

We shall not present a detailed  proof here because it is rather
large although straightforward; instead we give a short recipe for
it. To prove formula (\ref{formula}) it is appropriate to use
mathematical induction. First, the theorem is correct in the scalar
case ($n=0$). In this specific situation the formula is well known
(see, e.g., Ref.\,\cite{Vasiliev})
\begin{eqnarray}
\int^{\infty}_{-\infty}  \ldots \int^{\infty}_{-\infty}  \frac{d
{\bf k}^{(1)} \ldots d {\bf k}^{(l)} } {\left[v_{is}{\bf k}^{(i)}
\cdot {\bf k}^{(s)} + 2 {\bf a}^{(i)} \cdot
 {\bf k}^{(i)} +c  \right]^\alpha}
&=& \nonumber \\[3mm]
& & \hspace{-7.9cm} \frac{\pi^{\frac{d l}{2}}(\mathrm{det}
v)^{-\frac{d}{2}}\Gamma\left(\alpha -\frac{d l}{2}
\right)}{\Gamma(\alpha)} \left[ c - (v^{-1})_{is} {\bf a}^{(i)}
\cdot {\bf a}^{(s)} \right]^{
 \frac{d l}{2} - \alpha}\hspace{-0.8cm}. \label{scalar}
\end{eqnarray}
Now, let us suppose that formula (\ref{formula}) is valid for $n \in
\mathbb{N},\; n \geq 1$. Then if one differentiates both sides of
Eq.\,(\ref{formula}) with respect to $a^{(q_{n+1})}_{i_{n+1}}$
together with some cumbersome algebraic manipulations the formula
for $n+1$ is obtained.

\section{}\label{ApC}
In this appendix we present the integrals that were obtained during
the calculations of two-loop Feynman diagrams $B_1$ and $B_2$ which
are shown in Fig.\,\ref{fig3} (see Appendix\,\ref{ApA} for details
of the calculations). They are given in
Eqs.\,(\ref{int1})-(\ref{int3}). In addition we present here the
analytical expressions for integrals that were obtained by
comparison of our two-loop results for composite operators with
those obtained within the rapid-change model \cite{AdAnBaKaVa01}
[Eqs.\,(\ref{ApC4})-(\ref{ApC6})].

\begin{eqnarray}
\int_0^1\,d u_1\frac{u_1^{\frac{d}{2}-1}
\ln\left((1+u)(1-u_1)-\frac{(1-u_1)^2}{u_1+(1+u)(1-u_1)}\right)}{\left[u_1+(1+u)(1-u_1)\right]^{\frac{d}{2}+1}}
&=& \nonumber \\
& & \hspace{-8.5cm} \frac{4}{d(1+u)} \bigg\{
\frac{\psi(1)-\psi\left(d/2\right)}{2} -
\frac{{_2F_1}\Big(\frac{d}{2},\frac{d}{2};1+\frac{d}{2};\frac{u}{1+u}\Big)}{d(1+u)^{\frac{d}{2}}}
\nonumber \\
& &\hspace{-6.5cm} -
\frac{{_2F_1}\Big(1,1;2+\frac{d}{2};\frac{1}{(1+u)^2}\Big)}{(d+2)(1+u)^2}
\bigg\}, \label{int1}
\end{eqnarray}

\begin{eqnarray}
\int_0^1\,d u_1\frac{u_1^{\frac{d}{2}}
(1+u(1-u_1))^{-\frac{d}{2}-1}}{\left[u(2+u)-u_1(u^2+u-1)\right]}
&=& \nonumber \\
& & \hspace{-6.5cm} \frac{2
}{(d+2)(1+u)^2}\,{_2F_1}\Big(1,1;2+\frac{d}{2};\frac{1}{(1+u)^2}\Big),
\label{int2}
\end{eqnarray}

\begin{eqnarray}
\int_0^1 d u_1 \frac{u_1 [(u_1+(1+u)(1-u_1))(1+u)-(1-u_1)]
^{-\frac{1}{2}}}{(u_1+(1+u)(1-u_1))^2 (1-u_1)^{\frac{1}{2}}} & &
\nonumber \\ & & \hspace{-7.5cm}=
\frac{4}{3(1+u)}{_2F_1}\Big(\frac{1}{2},\frac{1}{2};\frac{5}{2};\frac{1}{(1+u)^2}\Big),
\label{int3}
\end{eqnarray}

\begin{eqnarray}
\int_{-1}^1 d x \,\,(1-x^2)^{d/2}\, x\,
\arctan\left(\frac{1+x}{\sqrt{1-x^2}}\right) && \nonumber \\ &&
\hspace{-5.5cm}= \frac12\frac{S_d}{S_{d-1}}\frac{d^2-1}{d(d+2)^2}\,,
\label{ApC4}
\end{eqnarray}

\begin{eqnarray}
\int_{-1}^1 d x \,\,\frac{(1-x^2)^{(d+1)/2}\, x}{\sqrt{4-x^2}}\,
\arctan\left(\frac{2+x}{\sqrt{4-x^2}}\right) && \nonumber \\ &&
\hspace{-7cm}= \frac18\frac{S_d}{S_{d-1}}\frac{d^2-1}{d(d+2)(d+4)}
{_2F_1}\Big(1,1;3+\frac{d}{2};\frac{1}{4}\Big) \,, \label{ApC5}
\end{eqnarray}

\begin{eqnarray}
\int_{-1}^1 d x \,\,\frac{(1-x^2)^{(d+1)/2}\, x^3}{\sqrt{4-x^2}}\,
\arctan\left(\frac{2+x}{\sqrt{4-x^2}}\right) && \nonumber \\ &&
\hspace{-7cm}= \frac18\frac{S_d}{S_{d-1}}\frac{d^2-1}{d(d+2)(d+4)}
 \nonumber \\ && \hspace{-7cm}
\times
\Biggl[{_2F_1}\Big(1,1;3+\frac{d}{2};\frac{1}{4}\Big)-\frac{d+3}{d+6}
\,\,{_2F_1}\Big(1,1;4+\frac{d}{2};\frac{1}{4}\Big)\Biggr] \,,
\nonumber \label{ApC6}
\end{eqnarray}

\section{}\label{ApD}

The explicit form of the coefficients $X_j^x$, $j=1,2,3,
x=K_1,K_{2a},\ldots,K_{2f}$ (see Sec.\ref{sec:CompOper}), in the MS
scheme is the following:
\begin{widetext}
\begin{equation}
X^{K_1}_1= \frac{S_d}{(2 \pi)^d} \left(\frac{\mu}{m}\right)^{2
\varepsilon} \frac{d-1}{4 u(1+u)}\frac{g}{\varepsilon}\,,\,\,\,\,\,
X^{K_1}_2 = 0\,,
\end{equation}
\begin{eqnarray}
X^{K_{2a}}_1&=& \frac{g^2 S_d^2}{(2\pi)^{2 d} 16 u^2
(1+u)(1-u^2)}\frac{d-1}{d}\frac{1}{\varepsilon}\left(\frac{\mu}{m}\right)^{4\varepsilon}
\Bigg[
\frac{{_2F_1}\left(1,1;2+\frac{d}{2};\frac{1}{(1+u)^2}\right)-u\,\,
{_2F_1}\left(1,1;2+\frac{d}{2};\frac{1}{2(1+u)}\right)}{(d+2)(1+u)}
\nonumber \\ && + \,\, \pi
\rho^2(d-2)\left(-\frac{1}{2}\,\,{_2F_1}\left(\frac{1}{2},\frac{1}{2};1+\frac{d}{2};\frac{1}{(1+u)^2}\right)
+\frac{u}{\sqrt{2(1+u)}}\,\,{_2F_1}\left(\frac{1}{2},\frac{1}{2};1+\frac{d}{2};\frac{1}{2(1+u)}\right)\right)
\Bigg]\,,\label{D2}
\end{eqnarray}
\begin{equation}
X^{K_{2a}}_2 = X^{K_{2a}}_{2a}+X^{K_{2a}}_{2b}\,,
\end{equation}
\begin{eqnarray}
X^{K_{2a}}_{2a} &=& \frac{g^2 S_d^2}{(2\pi)^{2 d} 32 u^3
(1+u)(1-u^2)}\frac{(d-1)(d+1)}{d(d+2)}\frac{1}{\varepsilon}\left(\frac{\mu}{m}\right)^{4\varepsilon}
\nonumber \\
&&\times
\left[(1+u)\,\,{_3F_2}\left(\frac{1}{2},1,1;\frac32,2+\frac{d}{2};1\right)-
{_3F_2}\left(\frac{1}{2},1,1;\frac32,2+\frac{d}{2};\frac{1}{(1+u)^2}\right)\right]\,,
\end{eqnarray}
\begin{eqnarray}
X^{K_{2a}}_{2b} &=& \frac{g^2 S_d S_{d-1}}{(2\pi)^{2 d} 16 u
(1+u)(1-u^2)}\frac{1}{\varepsilon}\left(\frac{\mu}{m}\right)^{4\varepsilon}
\int^1_{-1} dx \, \frac{(1-x^2)^{\frac{d+1}{2}}}{(1-u)^2+4 u x^2}
\nonumber \\
&& \left( -\frac{2 (1+u) x
\arctan\left(\frac{x}{\sqrt{1-x^2}}\right)}{\sqrt{1-x^2}} +
\frac{2(1+3 u) x
\arctan\left(\frac{x}{\sqrt{2(1+u)-x^2}}\right)}{\sqrt{2(1+u)-x^2}}
+ (u-1) \ln\left(\frac{2}{1+u}\right) \right)\,,
\end{eqnarray}
\begin{equation}
X^{K_{2b}}_1 = X^{K_{2b}}_{1a}+X^{K_{2b}}_{1b}\,,
\end{equation}
\begin{equation}
X^{K_{2b}}_{1a} = \frac{g^2 S_d^2}{(2\pi)^{2 d} 32 u^2
(1+u)^2}\frac{(d-1)^2}{d}\left(\frac{\mu}{m}\right)^{4\varepsilon}\frac{1}{\varepsilon}
\left(\frac{1}{\varepsilon}+
\frac{2\,\,{_2F_1}\left(1,1;2+\frac{d}{2};\frac{1}{(1+u)^2}\right)}{(d+2)(1+u)^2(1-u)}-\frac{u
\ln\left(\frac{1+u}{2}\right)}{(1-u)}\right)\,,
\end{equation}
\begin{eqnarray}
 X^{K_{2b}}_{1b} &=&
- \frac{g^2 S_d S_{d-1} (d-1)}{(2\pi)^{2d} 16 u
(1+u)(1-u^2)}\left(\frac{\mu}{m}\right)^{4\varepsilon}\frac{1}{\varepsilon}
\nonumber \\ &&\times  \int_{-1}^1 dx \, (1-x^2)^{\frac{d-1}{2}} x
\,\, \frac{\arctan\left(\frac{2+x}{\sqrt{2(1+u)-x^2}}\right)+
\arctan\left(\frac{1+u+x}{\sqrt{2(1+u)-x^2}}\right)}{\sqrt{2(1+u)-x^2}}\,,
\end{eqnarray}
\begin{equation}
X^{K_{2b}}_2 = X^{K_{2b}}_{2a}+X^{K_{2b}}_{2b}\,,
\end{equation}
\begin{eqnarray}
 X^{K_{2b}}_{2a}
&=& \frac{g^2 S_d^2}{(2\pi)^{2d} 32 u^2
(1+u)(1-u^2)}\frac{d^2-1}{d(d+2)}\left(\frac{\mu}{m}\right)^{4\varepsilon}\frac{1}{\varepsilon}
\Biggl\{ \frac{1-u}{\varepsilon}-1+\frac{4}{15(4+d)(6+d)u(1+u)^4}\nonumber \\
&&\times \Biggl[ 5(6+d)(1+u)^2\left((1+u)^3\,\,
{_3F_2}\left(1,1,\frac{3}{2};\frac52,3+\frac{d}{2};1\right)-
{_3F_2}\left(1,1,\frac32;\frac52,3+\frac{d}{2};\frac{1}{(1+u)^2}\right)\right)\nonumber \\
&& \hspace{5mm} + \left((1+u)^5\,\,
{_3F_2}\left(2,2,\frac{5}{2};\frac72,4+\frac{d}{2};1\right)-
{_3F_2}\left(2,2,\frac52;\frac72,4+\frac{d}{2};\frac{1}{(1+u)^2}\right)\right)
\Biggr] \Biggr\}\,,
\end{eqnarray}
\begin{eqnarray}
 X^{K_{2b}}_{2b} &=& - \frac{g^2 S_d S_{d-1}}{(2\pi)^{2 d} 16 u
(1+u)(1-u^2)}\left(\frac{\mu}{m}\right)^{4\varepsilon}\frac{1}{\varepsilon}
  \int_{-1}^1 dx \, \frac{(1-x^2)^{\frac{d+1}{2}}}{(1-u)^2+4 u x^2}
  \Biggl[(u-1+2 x^2)\ln\left(\frac{2}{1+u}\right) \\
  && \hspace{-15mm}+ 2 x \left(\frac{(1+u)(u-3+ 4 x^2)\arctan\left(\frac{1+x}{\sqrt{1-x^2}}\right)}{\sqrt{1-x^2}} +
  \frac{(3+u-2x^2)\left(\arctan\left(\frac{2+x}{\sqrt{2(1+u)-x^2}}\right)+
\arctan\left(\frac{1+u+x}{\sqrt{2(1+u)-x^2}}\right)\right)}{\sqrt{2(1+u)-x^2}}
\right) \Biggr]\,,\nonumber
\end{eqnarray}
\begin{eqnarray}
X^{K_{2c}}_1 &=& X^{K_{2c}}_{1a}+X^{K_{2c}}_{1b}\,, \\
X^{K_{2c}}_{1a} &=& -\frac{1}{1+u} X^{K_{2b}}_{1a}\,,\\
X^{K_{2c}}_{1b} &=& -\frac{1}{2} X^{K_{2b}}_{1b}\,,\\
X^{K_{2c}}_{2} &=&0\,,
\end{eqnarray}
\begin{eqnarray}
X^{K_{2d}}_1 &=&- \frac{g^2 S_d^2}{(2\pi)^{2d} 16 u^2
(1+u)(1-u^2)}\frac{d-1}{d}\left(\frac{\mu}{m}\right)^{4\varepsilon}\frac{1}{\varepsilon}\nonumber
\\ &&\times \Biggl\{ \frac{1}{(2+d)(1+u)}
\left[\frac{{_2F_1}\left(1,1;2+\frac{d}{2};\frac{1}{(1+u)^2}\right)}{1+u}-
\frac{u\,\,{_2F_1}\left(1,1;2+\frac{d}{2};\frac{1}{2(1+u)}\right)}{2}\right] \nonumber \\
&& \hspace{5mm}+ \frac{(d-2)\pi
\rho^2}{2}\left[-\frac{{_2F_1}\left(\frac12,\frac12;1+\frac{d}{2};\frac{1}{(1+u)^2}\right)}{1+u}+
\frac{u\,\,{_2F_1}\left(\frac12,\frac12;1+\frac{d}{2};\frac{1}{2(1+u)}\right)}{\sqrt{2(1+u)}}\right]\Biggr\}\,,\label{D16}
\\ X^{K_{2d}}_2&=&0\,,
\end{eqnarray}
\begin{equation}
X^{K_{2e}}_1 = X^{K_{2e}}_{1a}+X^{K_{2e}}_{1b}\,,
\end{equation}
\begin{eqnarray}
X^{K_{2e}}_{1a}&=& \frac19 \frac{g^2 S_d^2}{(2\pi)^{2d} 32 u^2
(1+u)^2(1-u)^2}\frac{d-1}{d(d+2)}\left(\frac{\mu}{m}\right)^{4\varepsilon}\frac{1}{\varepsilon}
 \Biggl\{ \frac{1}{\varepsilon} (d+3)(d-2)(1-u)^2
 + \frac{1}{2(2+d)^2(1+u)^2} \nonumber \\
 && \times \Biggl[
 4(d+2)(d+1)u\left(3u(1+u)^2\,\,{_2F_1}\left(1,1;1+\frac{d}{2};\frac{1}{4}\right)+
 4(2+u)\,\,{_2F_1}\left(1,1;1+\frac{d}{2};\frac{1}{(1+u)^2}\right)\right)
 \nonumber \\
&& \hspace{5mm} + u^2(1+u)^2(d^3-9d^2-48d-44)
{_2F_1}\left(1,1;2+\frac{d}{2};\frac{1}{4}\right) \nonumber \\
&& \hspace{5mm} +
4\left(d^3+d^2(3-8u-4u^2)-12d(1+u)^2-4(5+4u+2u^2)\right)\,\,{_2F_1}\left(1,1;2+\frac{d}{2};\frac{1}{(1+u)^2}\right)
 \Biggr]\nonumber \\
 &&-\frac{2(d-2)}{1+u}\left(u^2(1+u)\,\,{_2F_1}\left(1,1;2+\frac{d}{2};\frac{1}{4}\right)+
 2\,\,{_2F_1}\left(1,1;2+\frac{d}{2};\frac{1}{(1+u)^2}\right)\right)
 \nonumber \\
 &&+ \frac{2u(3+u)(d-2)}{1+u}\,\, {_2F_1}\left(1,1;2+\frac{d}{2};\frac{1}{2(1+u)}\right)
 \Biggr\}\,,
\end{eqnarray}
\begin{eqnarray}
X^{K_{2e}}_{1b} &=& \frac19 \frac{g^2 S_d S_{d-1}}{(2\pi)^{2 d} 8 u
(1+u)^2(1-u)^2}\left(\frac{\mu}{m}\right)^{4\varepsilon}\frac{1}{\varepsilon}
\nonumber \\
&& \times
  \int_{-1}^1 dx \, (1-x^2)^{\frac{d-1}{2}}(1-d+4 x^2)\, x\,
\frac{\arctan\left(\frac{2+x}{\sqrt{2(1+u)-x^2}}\right)+
\arctan\left(\frac{1+u+x}{\sqrt{2(1+u)-x^2}}\right)}{\sqrt{2(1+u)-x^2}}\,,
\end{eqnarray}
\begin{equation}
X^{K_{2e}}_2 = X^{K_{2e}}_{2a}+X^{K_{2e}}_{2b}\,,
\end{equation}
\begin{eqnarray}
X^{K_{2e}}_{2a} &=& \frac16 \frac{g^2 S_d^2}{(2\pi)^{2d} 16 u^2
(1+u)^2(1-u)^2}\frac{d^2-1}{d(d+2)}\left(\frac{\mu}{m}\right)^{4\varepsilon}\frac{1}{\varepsilon}
\Biggl\{ \frac{1}{\varepsilon} (1-u)^2
 + \frac{1}{2(2+d)^2(1+u)^2} \nonumber \\
 && \times \Biggl[ u(d+2)\left(3u(1+u)^2\,\,{_2F_1}\left(1,1;1+\frac{d}{2};\frac{1}{4}\right)
 + 4(2+u) {_2F_1}\left(1,1;1+\frac{d}{2};\frac{1}{(1+u)^2}\right) \right)  \\
 && \hspace{-7mm} -2u^2(d+3)(1+u)^2\,\,{_2F_1}\left(1,1;2+\frac{d}{2};\frac{1}{4}\right)-
 4(2u(2+u)+d(2u-1+u^2))\,\,{_2F_1}\left(1,1;2+\frac{d}{2};\frac{1}{(1+u)^2}\right)
 \Biggr] \Biggr\}\,,\nonumber
\end{eqnarray}
\begin{eqnarray}
X^{K_{2e}}_{2b} &=& -\frac16 \frac{g^2 S_d S_{d-1}}{(2\pi)^{2 d} 4 u
(1+u)^2(1-u)^2}\left(\frac{\mu}{m}\right)^{4\varepsilon}\frac{1}{\varepsilon}
\nonumber \\
&& \times
  \int_{-1}^1 dx \, (1-x^2)^{\frac{d+1}{2}}\, x\,
\frac{\arctan\left(\frac{2+x}{\sqrt{2(1+u)-x^2}}\right)+
\arctan\left(\frac{1+u+x}{\sqrt{2(1+u)-x^2}}\right)}{\sqrt{2(1+u)-x^2}}\,,
\end{eqnarray}
\begin{eqnarray}
X^{K_{2f}}_{1} &=& \frac19 \frac{g^2 S_d^2}{(2\pi)^{2 d} 16 u^2
(1+u)^2}
\left(\frac{\mu}{m}\right)^{4\varepsilon}\frac{1}{\varepsilon^2}
\frac{d^4+4d^3+d^2-10d+4}{d(d+2)}\,, \\
X^{K_{2f}}_{2} &=& \frac{1}{72} \frac{g^2 S_d^2}{(2\pi)^{2 d} 16 u^2
(1+u)^2}
\left(\frac{\mu}{m}\right)^{4\varepsilon}\frac{1}{\varepsilon^2}
\frac{4(6-7d+d^3)}{d(d+2)}\,, \\
X^{K_{2f}}_{3} &=& \frac{1}{24} \frac{g^2 S_d^2}{(2\pi)^{2 d} 16 u^2
(1+u)^2}
\left(\frac{\mu}{m}\right)^{4\varepsilon}\frac{1}{\varepsilon^2}
\frac{4(d^2-1)}{d(d+2)}\,,
\end{eqnarray}
\end{widetext}

\end{document}